\documentclass[pra,aps,floatfix,10pt,twocolumn,nofootinbib,superscriptaddress]{revtex4-2} 

\usepackage[T1]{fontenc}
\usepackage{lmodern}
\usepackage[utf8]{inputenc}
\usepackage{microtype}
\usepackage{array}
\usepackage{graphicx}
\usepackage{dcolumn}
\usepackage{bm}
\usepackage{amsmath}
\usepackage{amsfonts}
\usepackage{amssymb}
\usepackage{amstext}   
\usepackage{amsthm}
\usepackage{array}
\usepackage[english]{babel}
\usepackage{makecell}
\usepackage{dsfont}
\usepackage{nameref}
\usepackage{color}
\usepackage[outdir=./]{epstopdf}
\usepackage[colorlinks = true, citecolor = blue, urlcolor =cyan]{hyperref}
\usepackage[figure]{hypcap}
\usepackage{multirow}
\usepackage{makecell}
\usepackage{xcolor}
\usepackage{mathtools}
\usepackage{mathdots}
\usepackage[notransparent]{svg}
\newcolumntype{C}{>{$}c<{$}}
\AtBeginDocument{
\heavyrulewidth=.08em
\lightrulewidth=.05em
\cmidrulewidth=.03em
\belowrulesep=.65ex
\belowbottomsep=0pt
\aboverulesep=.4ex
\abovetopsep=0pt
\cmidrulesep=\doublerulesep
\cmidrulekern=.5em
\defaultaddspace=.5em
\tabcolsep=7pt
}
\usepackage{booktabs}
\usepackage{soul}

\definecolor{emerald}{rgb}{0.07, 0.53, 0.03}

\usepackage{braket}

\usepackage{amsbsy}

\usepackage{nicefrac} 

\usepackage{soul}   
\setstcolor{red} 
\setul{}{1.5pt}  

\begin{document}

\title{A strong-driving toolkit for topological Floquet energy pumps with superconducting circuits}
\author{Martin Ritter}
\affiliation{Joint Quantum Institute, Department of Physics, University of Maryland, College Park, Maryland 20742, USA}

\author{David M. Long}
\affiliation{Joint Quantum Institute, Department of Physics, University of Maryland, College Park, Maryland 20742, USA}
\affiliation{Condensed Matter Theory Center, Department of Physics, University of Maryland, College Park, Maryland 20742, USA}
\affiliation{Department of Physics, Stanford University, Stanford, California 94305, USA}

\author{Qianao Yue}
\affiliation{Joint Quantum Institute, Department of Physics, University of Maryland, College Park, Maryland 20742, USA}

\author{Maya Amouzegar}
\affiliation{Joint Quantum Institute, Department of Physics, University of Maryland, College Park, Maryland 20742, USA}

\author{Anushya Chandran}
\affiliation{Department of Physics, Boston University, 590 Commonwealth Avenue, Boston, Massachusetts 02215, USA}
\affiliation{Max-Planck-Institut f\"{u}r Physik komplexer Systeme, N\"othnitzer Str. 38, 01187 Dresden, Germany}

\author{Alicia J.  Koll\'ar}
\affiliation{Joint Quantum Institute, Department of Physics, University of Maryland, College Park, Maryland 20742, USA}

\preprint{APS/123-QED}

\date{\today}

\begin{abstract}

Topological Floquet energy pumps---which use periodic driving to create a topologically protected quantized energy current---have been proposed and studied theoretically, but have never been observed directly. 
Previous work~\cite{Long2022b} proposed that such a pump could be realized with a strongly-driven superconducting qubit coupled to a cavity.
Here, we experimentally demonstrate that the proposed hierarchy of energy scales and drive frequencies can be realized using a transmon qubit. 
We develop an experimental toolkit to realize the adiabatic driving field required for energy pumping using coordinated frequency modulation of the transmon and amplitude modulation of an applied resonant microwave drive.
With this toolkit, we measure adiabatic evolution of the qubit under the applied field for times comparable to $T_1$, which far exceed the bare qubit dephasing time.
This result paves the way for direct experimental observation of topological energy pumping.

\end{abstract}
\maketitle

\section{Introduction}\label{sec:intro}

\begin{figure}[ht]
    \centering
    \includegraphics[width=\linewidth]{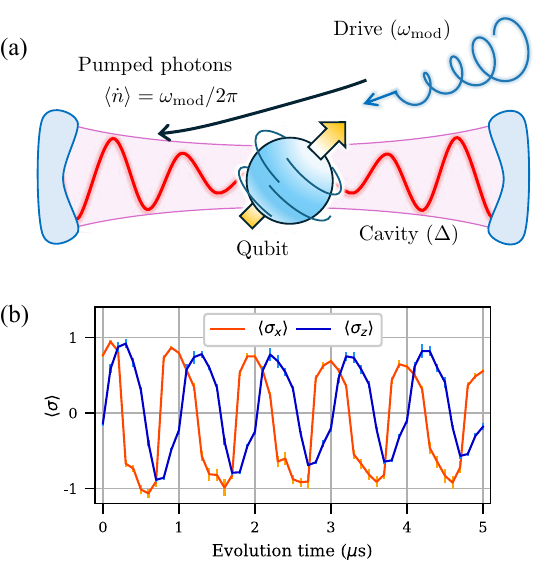}
    \caption{The topological energy pump. (a) A qubit is driven with two oscillatory fields: an external drive (blue) and a cavity mode (red). In the topological regime, the qubit absorbs energy from the drive and pumps energy into the cavity (the direction of the pump can be reversed by flipping the state of the qubit). For the experimental implementation, the external drive is a combination of a static $x$ field and a circularly polarized field in the $xz$ plane. (b) Measurement of the qubit following the rotating field from part (a). The qubit adiabatically follows the applied field as seen by the 90-degree phase shift between the $x$ and $z$ projections of the state vector.}
    \label{fig:pumping_schematic}
\end{figure}

The adiabatic evolution of a system, in which the parameters of a non-degenerate Hamiltonian are varied slowly compared to gaps between energy levels~\cite{Born1928, general_adiabatic_review}, provides a general protocol for state preparation, as well as a quantum computation paradigm~\cite{adiabatic_QC_Farhi_2001,adiabatic_geometry_review_Kolodrubetz_2017, adiabaticQC_review_Albash_2018}. 
Cyclic adiabatic modulation of the Hamiltonian parameters, i.e. when the path in Hamiltonian parameter space is closed, can generate \emph{pumps} in which a net change in a system observable occurs after every cycle, even though the eigenspectrum returns to itself~\cite{adiabatic_review_periodically_drive_Weinberg_2017}.
 
The topological pumps studied in this article use periodic driving to generate an average quantized current across a system, where the quantization is set by a topological invariant of the system \cite{Thouless1983pump,Citro:2023_thoulesspumpreview,Martin:2017aa,Crowley:2019_classification,Nathan:2019_drivendiss,PSAROUDAKI2021,Long2022b,Long2021:class,Nathan2020b, schroer_2014_topo_transition_spin_12,Sridhar2024photonic,Koch2024dissipative}. 
These were first explored in the context of the Thouless pump~\cite{Thouless1983pump}: an adiabatically modulated insulating wire that pumps integer units of charge across the wire every driving period. 
A similar pump applied to a ladder of photon number states, rather than a wire, pumps \emph{energy} into or out of a cavity at a topologically quantized rate \cite{Martin:2017aa,Crowley:2019_classification,Nathan:2019_drivendiss,Long2022b}.
This pump can be implemented using a cavity coupled to a qubit which is subject to a strong elliptically polarized external drive, as shown schematically in Fig.~\ref{fig:pumping_schematic}(a).
The resulting energy current could be used to rapidly cool cavities to their ground states \cite{Nathan:2019_drivendiss} or to prepare large non-classical states such as Fock states \cite{Long2022b}.
Other implementations, not using a qubit, have been proposed in Refs.~\cite{Yuan2018ringresonator,Nathan2022weyl,Lantagne-Hurtubise2024graphene,Luneau2022power,Long2021:class,Nathan2020b,Schwennicke2022enantioselective}.

While experiments have measured the quantized transport of particles for a Thouless pump~\cite{Cheng_topo_pump_metamaterial_2020,quasi_crystal_photonic_Thouless, fermion_Thouless_pump, Lohse2018, Thouless_pump_Lohse_2016, Thouless_pump_disordered_photonics_Cerjan,Jurgensen_nonlinear_Thouless_pump_2021,Mostaan_nonlinear_Thouless_soliton_atoms_2022,grinberg_robust_2020, walter_quantization_2023, nakajima_competition_2021, jurgensen_quantized_2023,SC_Thouless_2025}, no direct measurements of the energy current have been performed for the driven qubit-cavity system. 
Nitrogen-vacancy centers in diamond~\cite{Boyers2020} and superconducting qubits~\cite{cQED_two_classical_drives_Malz} have been used to study the limit of large photon numbers, in which the cavity can be treated as a classical drive.
However, in this regime, the topological current is overwhelmed by photon number fluctuations in the drive and cannot be observed directly.

Directly measuring the quantized energy current thus requires realizing the topological regime at low photon number in a low-loss cavity.
Remaining adiabatic at small photon numbers further necessitates (i) a large single-photon coupling, and (ii) a strong external field. 
The circuit QED platform of superconducting qubits coupled to microwave cavities is ideal for satisfying these two conditions~\cite{Blais_review, quantum_engineer_guide}, and Ref.~\cite{Long2022b} proposed that the energy pump can be achieved in a rotating frame for realistic device parameters.

In this article, we address the strong external drive component of the proposed topological energy pump and show experimentally that the required field strengths for the energy pump can be reached in a transmon qubit. 
Our implementation uses simultaneous modulation of both the qubit frequency and the amplitude of a microwave drive in order to generate a circularly-polarized effective field rotating in the $xz$-plane of the Bloch sphere of the qubit. 
We develop a robust set of characterization measurements for both field components, including a fast measurement to characterize the delay between them. 
We demonstrate that large effective field strengths exceeding 100 MHz can be synthesized on a transmon qubit without being negatively impacted by the higher energy levels of the transmon.
Finally, using measurement techniques developed in Ref.~\cite{dissipation_paper}, we show that the qubit adiabatically follows the applied effective field [Fig.~\ref{fig:pumping_schematic}(b)] for a time which far exceeds the native qubit phase coherence times. 

Our results show that the hardware requirements for the topological energy pump can be realized in a single circuit-QED device and that the performance of the pump should not be limited by the short phase coherence times that plague flux-tunable qubit architectures.

The rest of the manuscript is organized as follows. In Sec. \ref{sec:topo_pump}, we provide a brief introduction to the topological pump and review the required energy hierarchy. Sec. \ref{sec:exp_implementation} describes the experimental implementation and relevant energy scales. In Sec. \ref{sec:b_cal}, we characterize the drive field components and 
demonstrate synchronization
between them, implementing a circularly-polarized drive. Finally, in Sec.~\ref{sec:following}, we present experimental data demonstrating that the qubit adiabatically follows the drive field and provide limits on the drive frequency and strength needed to guarantee adiabaticity.

\section{Driven systems with synthetic topology}
\label{sec:topo_pump}

The experimental toolkit developed in this work will make it possible to realize and directly observe a \emph{topological energy pump}~\cite{Martin:2017aa,Kolodrubetz2018,Crowley:2019_classification,Nathan:2019_drivendiss,Long2021:class,Nathan2020b}. In this section, we review the theoretical description of these pumps, and establish the experimental requirements that must be met in order to observe energy pumping.

Topological energy pumps (also called topological photon pumps or topological frequency converters) were first proposed in qubit systems with strong, adiabatic (slow) driving by classical fields~\cite{Martin:2017aa,Crowley:2019_classification,Crowley:2020tl,Boyers2020,cQED_two_classical_drives_Malz,Long2022b}. The Hamiltonian for this system is 
\begin{equation}\label{eqn:Hmodel}
    H_B = \frac{1}{2}\vec{\sigma}\cdot\vec{B}(t),
\end{equation}
where \(\vec{\sigma}\) is a vector of qubit Pauli operators, and \(\vec{B}(t)\) is the vector of their time-dependent coefficients. We refer to \(\vec{B}(t)\) as a magnetic field, regardless of whether it is implemented through the coupling of a real magnetic field to the magnetic moment of a spin-\(1/2\) particle.

When the field \(\vec{B}(t)\) is a superposition of two orthogonally oriented circularly polarized drives and a static field, the qubit can preferentially absorb energy from one driving field and emit into the other field~\cite{Martin:2017aa}. Anticipating the model we realize experimentally, we say the qubit absorbs from a drive with (angular) frequency \(\omega_{\mathrm{mod}} = 2\pi/T_{\mathrm{mod}}\) and emits into a drive with frequency \(\Delta\). The average power pumped by the qubit is quantized when \(\omega_{\mathrm{mod}}/\Delta \not\in \mathbb{Q}\) is irrational, even though there is no exactly resonant conversion process between quanta of energy \(\omega_{\mathrm{mod}}\) and \(\Delta\)~\cite{Martin:2017aa}.

It was recognized in Ref.~\cite{Nathan:2019_drivendiss} that the same phenomenon occurs when one (or both) of the classical drives is replaced by the quantized mode of an oscillator, and in Ref.~\cite{Long2022b} it was shown that the topological pumping mechanism can even \emph{boost} non-classical states of a cavity. That is, an initial unentangled state of the qubit and cavity of the form
\begin{equation}
	\ket{\psi(0)} = \ket{s}\otimes\sum_{n} c_n \ket{n}
\end{equation}
may be subjected to a time-periodic drive on the qubit which, at special times \(T_N\), produces a time-evolved state
\begin{equation}
	\ket{\psi(T_N)} \approx \ket{s}\otimes\sum_{n} c_n \ket{n+ h_N}.
	\label{eqn:boost}
\end{equation}
Above, $h_N$ is an integer equal to the number of periods of the drive $T_N\omega_{\mathrm{mod}}/2\pi = T_N/T_{\mathrm{mod}}$. The topological pump acts to translate the cavity state in Fock space. This can, for instance, enable the preparation of highly excited Fock states provided that Fock states with a smaller number of photons can be reliably prepared. Such highly excited Fock states are a useful resource for quantum metrology~\cite{Giovannetti2011metrology}.

A Hamiltonian which achieves topological energy pumping between a classical drive and a quantum cavity with annihilation operator \(a\) is~\cite{Nathan:2019_drivendiss,Long2022b}
\begin{equation}
    H_{B}^+ = \frac{1}{2}\vec{\sigma}\cdot\vec{B}(t) + \Delta\, a^\dagger a + g\,(a^\dagger \sigma^- + a \sigma^+),
    \label{eqn:Hamiltonian}
\end{equation}
where
\begin{equation}
    \vec{\sigma}\cdot\vec{B}(t) = B_0 \{  \sin(\omega_{\mathrm{mod}} t) \sigma_z +
    [m+\cos(\omega_{\mathrm{mod}} t)] \sigma_x \}
    \label{eqn:Bfield}
\end{equation}
is a periodic drive on the qubit, \(m\) is a dimensionless parameter, \(\sigma_{x,y,z}\) are Pauli matrices, and \(\sigma^\pm = (\sigma_x \pm i \sigma_y)/2\). 

There is a simple picture for topological pumping of Fock states with the Hamiltonian in Eq.~\eqref{eqn:Hamiltonian} when we set \(m=1\) and neglect evolution of the cavity when it is not resonant with the qubit. At the beginning of the protocol (\(t=0\)) the cavity is initialized in an accessible Fock state \(\ket{n}\) and the qubit is prepared in the instantaneous excited state \(\ket{+}\) of \(\vec{\sigma} \cdot \vec{B}(0)\), which is aligned in the \(x\) direction on the Bloch sphere if \(B_0 > 0\). The qubit and cavity begin far off-resonant, and the qubit adiabatically follows the rotating external field \(\vec{B}(t)\) until \(t \approx T_{\mathrm{mod}}/2\), by which time the qubit is in the state aligned along $+z$, denoted by $\ket{e}$. Near \(t \approx T_{\mathrm{mod}}/2\), the external field functions as a ramp, slowly driving the qubit through an avoided level crossing with the cavity, causing an adiabatic transition from the \(\ket{e,n}\) state to the state \(\ket{g,n+1}\) (where \(\ket{g}\) is the state aligned along \(-z\)) as the cavity absorbs a qubit excitation. At the end of this ramp, the field on the qubit is aligned along the \(-z\) direction, so that the qubit is again in an excited state of \(\vec{\sigma} \cdot \vec{B}\). The external field then slowly returns to its large initial value, with the qubit adiabatically following the field and ending the Floquet period in an excited state of \(\vec{\sigma} \cdot \vec{B}(T_{\mathrm{mod}}) = \vec{\sigma} \cdot \vec{B}(0)\). The state of the qubit thus returns to its initial condition, except that the cavity now has \(n+1\) photons. The protocol is then repeated at a fixed rate \(1/T_\mathrm{mod}\), pumping an average of one photon per period into the cavity.

More generally, topological pumping can be understood as Thouless pumping~\cite{Thouless1983pump,Citro2023pumpreview} in the Fock basis. This is a topological effect which requires adiabaticity of the drives and the existence of a nonvanishing Chern number \(C\) in the Fock-basis tight-binding model, which is effectively extended by an additional \emph{synthetic dimension} due to the periodic drive~\cite{Ozawa2019syntheticreview,Sambe1973synthetic,Ho1983,Verdeny_2016_synthetic_dim,Martin:2017aa,Crowley:2019_classification}. In the toy picture, a non-zero \(C\) simply verifies that the external field passes close enough to zero for the qubit to exchange excitations with the cavity [\(C=\pm 1\) in the model of Eq.~\eqref{eqn:Hamiltonian}]. The sign of the Chern number indicates whether the state loses an excitation to the cavity, or absorbs an excitation from the cavity, so that the average photon pumping rate is
\begin{equation}
    \lim_{t\to\infty}\frac{1}{t}\int_0^t \langle\dot{n}(t')\rangle \mathrm{d}t' = \frac{C}{T_{\mathrm{mod}}}.
    \label{eqn:pump_rate}
\end{equation}

In order to realize pumping, we must operate in the adiabatic regime, which implies some restrictions on experimental parameters. The frequencies \(\omega_{\mathrm{mod}}\) and \(\Delta\) must be much smaller than the external field strength \(B_0\) and the qubit-cavity coupling \(g \sqrt{n}\). In the cavity or circuit QED context, it is most natural to realize this requirement in a rotating frame, so that \(\Delta\) is a detuning between a large lab-frame cavity frequency, \(\omega_c\), and the frequency of the rotating frame, \(\omega_d\). That is, \(\Delta = \omega_c - \omega_d\). Similarly, the mean rotating frame qubit frequency is given by \(\omega_q- \omega_d\), where \(\omega_q\) is the mean lab-frame qubit frequency. As Eq.~\eqref{eqn:Hamiltonian} has a mean qubit frequency of zero, we set \(\omega_d = \omega_q\). The Hamiltonian~\eqref{eqn:Hamiltonian} would then come with counter-rotating terms of frequency \(\omega_q + \omega_c\), which can be ignored when this frequency is large compared to other scales. Our experiment uses a transmon qubit~\cite{Koch2007transmon,Schreier2008transmon}, which has a relatively low anharmonicity \(\alpha\). In order to treat the transmon as having two levels, we need that this \(\alpha\) is also much larger than any Hamiltonian parameter in Eq.~\eqref{eqn:Hamiltonian}, which is a much stronger constraint than that imposed by \(\omega_q + \omega_c\). In summary, operating in the adiabatic regime requires that
\begin{equation}
    \frac{1}{T_{\mathrm{coh}}} \ll\Delta, \omega_{\mathrm{mod}} \ll B_0, g \sqrt{n} \ll \alpha, \omega_q+\omega_c,
    \label{eqn:operating_regime}
\end{equation}
where $T_{\mathrm{coh}}$ is the coherence time of the system and sets the lower limit for $\Delta$ and  $\omega_{\mathrm{mod}}$. Furthermore, the requirement of non-vanishing Chern number imposes~\cite{Nathan:2019_drivendiss}
\begin{equation}
    B_0^2 (1-m)^2 < g^2 n < B_0^2(1 + m)^2.
    \label{eqn:topo_regime}
\end{equation}
We see that the photon occupation number \(n\) cannot be too small or large for pumping to occur.
We refer to the range of parameters specified by  Eq.~\eqref{eqn:topo_regime} as the \emph{topological regime}.

In the remainder of the article, we develop the experimental tools to engineer the Hamiltonian~\eqref{eqn:Hamiltonian} in the adiabatic regime. We describe how to generate a strong, coherent, and slowly varying field $\vec{B}(t)$ in the rotating frame satisfying $B_0\gg g$, and verify adiabatic following of the qubit in the absence of the cavity. The relevant coherence time for the system is approximately the qubit lifetime, $T_1$, and is not limited by the relatively short dephasing time, $T_2$.  

\begin{figure*}
    \centering
    \includegraphics[width=\linewidth]{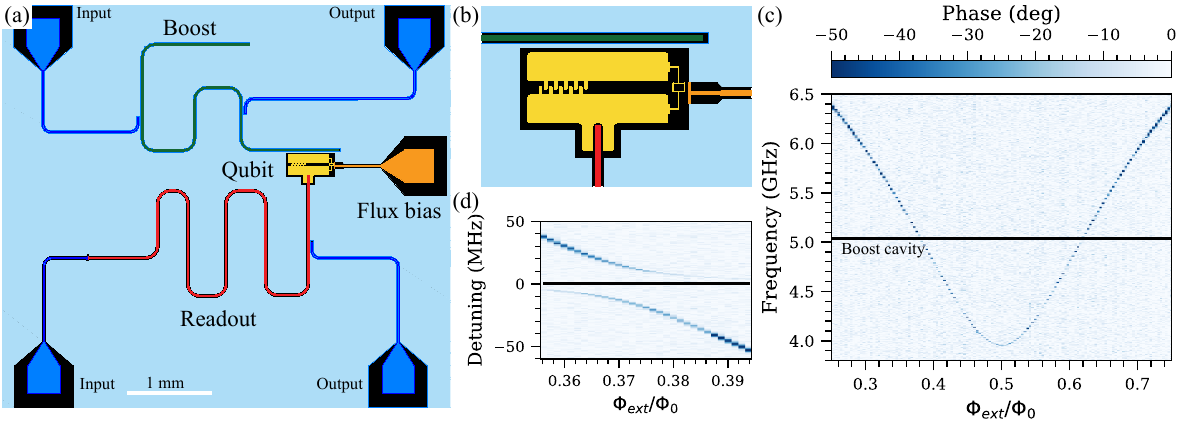}
    \caption{Device layout and qubit spectroscopy. (a) Full device layout: readout cavity in red, tunable transmon in yellow with associated flux bias line (orange), a low-loss boost cavity used for the full topological pumping protocol (green, unused in this experiment), and an external magnet for DC bias (not shown). (b) Zoom in on the qubit layout. (c) Qubit spectroscopy as a function of externally applied flux (in units of $\Phi_0$, the flux quantum). The qubit frequency is measured using dispersive readout and the colorbar indicates the phase shift on the readout cavity. The qubit is tunable over a 2.5 GHz range and can be brought into resonance with the boost cavity (black line at 5.04 GHz). The resulting avoided crossing is shown in (d) where the detuning is measured relative to the boost cavity frequency.}
    \label{fig:device_and_spec}
\end{figure*}

\section{Experimental Implementation and Device Design}\label{sec:exp_implementation}
In this section, we show that the qubit portion of the topological energy pump Hamiltonian [Eq.~\eqref{eqn:Hamiltonian}] can be realized in the circuit QED platform, and present a device which satisfies the required hierarchy of scales shown in Eq.~\eqref{eqn:operating_regime}.

Our experimental platform consists of a driven tunable transmon coupled to a cavity and is described using the Jaynes-Cummings Hamiltonian \cite{wallsMilburn, Blais_review, quantum_engineer_guide}:
\begin{multline}\label{eq:JC}
    H = \omega_c a^{\dagger}a +\omega_q \frac{\sigma_z}{2}+g(a^{\dagger}\sigma^-+a \sigma^+)\\ 
    +\Omega \cos(\omega_d t)\sigma_x
\end{multline}
where $\omega_{c,q,d}$ are the frequencies of the cavity, qubit, and drive respectively; $a$ is the bosonic annihilation operator for the cavity; $\sigma^-$ is the lowering operator for the qubit; and $\Omega, g$ are the driving strength and qubit-cavity coupling strength, respectively. 
The device design, shown in Fig.~\ref{fig:device_and_spec}(a), features a qubit coupled to two cavities. One cavity, which we denote as the boost cavity, is a cavity with high quality factor (Q) configured to host the topological pump, whereas a second, lower Q, cavity is used to read out the state of the transmon. This readout cavity is also used to apply drives to the qubit. 
The qubit frequency is tuned through the application of an external magnetic flux either from an on-chip flux bias line or from an external magnet. 

We synthesize the required rotating field in the \emph{rotating frame} of the applied qubit drive [specified by the transformation $U=\exp{(i\omega_d t[\hat{n}+\frac{\sigma_z}{2}])}$]. Rewriting the Jaynes-Cummings Hamiltonian in this frame and dropping rapidly oscillating terms we obtain:
\begin{equation}
    H = \Delta a^{\dagger}a + \delta \frac{\sigma_z}{2}+g(a^{\dagger}\sigma^-+a \sigma^+)+\Omega\frac{\sigma_x}{2}
\end{equation}
where $\Delta=\omega_c-\omega_d$ ($\delta=\omega_q-\omega_d$) is the detuning between the cavity (qubit) and the applied drive. 
By comparing terms to Eq.~\eqref{eqn:Hamiltonian}, we obtain the required form for $\delta$ and $\Omega$ to synthesize the topological pump Hamiltonian:
\begin{equation}
    \delta(t) = B_0 \sin(\omega_{\mathrm{mod}}t), ~
    \Omega(t) = B_0 [m+\cos(\omega_{\mathrm{mod}}t)]. 
\end{equation}
Above, $B_0$ and $\omega_{\mathrm{mod}}$ are the amplitude and frequency of the effective magnetic field respectively. We refer to the $\delta$ term as an effective $z$-magnetic field and the $\Omega$ term as an effective $x$-magnetic field. Their generation and calibration are described in Section \ref{sec:b_cal}.

The readout cavity [red trace in Fig.~\ref{fig:device_and_spec}(a)] is a coplanar waveguide half-wave resonator with resonant frequency 7.492~GHz and linewidth 350~kHz. As the qubit microwave drives are also applied through this cavity, it is configured in a transmission geometry to reduce the power reaching the readout amplification chain.
The cavity is capacitively coupled to the qubit with coupling strength 90 MHz to facilitate large microwave drive strengths on the qubit. 

The tunable transmon [yellow in Fig.~\ref{fig:device_and_spec}(a) and (b)] consists of a DC SQUID loop in parallel with floating capacitor pads. The SQUID loop is made with asymmetric junctions (with an inductive energy ratio $E_{J_1}=3E_{J_2}$), reducing the noise from the flux bias line by decreasing the frequency tuning range. A relatively large charging energy of 240 MHz is used for this qubit to improve the performance at large microwave drive strengths. Qubit spectroscopy as a function of applied external flux is performed using two-tone spectroscopy \cite{Blais_review} showing a tuning range from 4 to 6.4 GHz in Fig.~\ref{fig:device_and_spec}(c). The qubit can be tuned into resonance with the boost cavity. The resulting avoided crossing is shown in Fig.~\ref{fig:device_and_spec}(d), from which the qubit-boost coupling strength is extracted (\(g = 13\)~MHz). 

The qubit frequency is controlled by two sources of external flux: an external magnet providing the DC operating frequency of the qubit and an on-chip flux bias line [orange in Fig.~\ref{fig:device_and_spec}(a)] with a 5 MHz bandwidth for AC modulation. The on-chip flux bias line (FBL) is driven with an arbitrary waveform generator (AWG) to provide full control over the frequency modulation. 

The boost cavity [green trace in Fig.~\ref{fig:device_and_spec}(a)] consists of a quarter-wave coplanar waveguide resonator at 5.04 GHz with power loss rate $\kappa=84$~kHz. It is designed to store the photonic state in the full implementation of the topological pump described in the introduction, but is unused in this manuscript. The particular 90-degree orientation of the readout and boost cavities minimizes the direct classical crosstalk between the cavities by introducing an effective mode-mismatch. 

\section{B Field Components}\label{sec:b_cal}
\begin{figure}
    \centering
    \includegraphics[width=\linewidth]{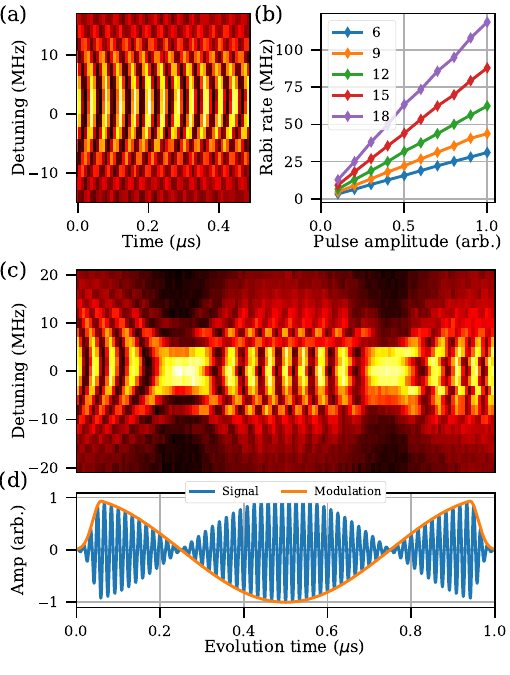}
    \caption{Calibration of $B_x$: (a) Rabi oscillations as a function of qubit-drive detuning and evolution time showing the usual Rabi chevron pattern. The color indicates the probability of the qubit being in the excited state with yellow being $\ket{e}$ and black being $\ket{g}$. (b) Extracted Rabi rate as a function of AWG modulation amplitude for different generator output powers (legend in units of dBm), achieving a peak Rabi rate of 110 MHz. (c) Rabi chevron for amplitude modulated drive at $\omega_{\mathrm{mod}}=1$ MHz [modulating waveform shown in (d)]. The amplitude of the drive is directly mapped to the qubit population oscillation frequency as seen by the fact that the zero crossings in (d) are accompanied by a freezing of the qubit dynamics. For clarity, we present an example waveform with a much lower carrier frequency in (d). 
    }
    \label{fig:bx_cal}
\end{figure}


Achieving the Hamiltonian in Eq.~\eqref{eqn:Hamiltonian} requires a time-dependent effective magnetic field in the $x$-direction ($B_x$) and $z$-direction ($B_z$).
In the section below, we present the experimental procedure to implement and calibrate $B_x$ and $B_z$ respectively. As the field components originate from different sources, we also develop a method to measure the delay between them. In particular, we synthesize a \emph{circularly} polarized field of the form:
\begin{equation}
    \vec{B} = B_0 \begin{pmatrix} \cos(\omega_{\mathrm{mod}}t) \\0\\\sin(\omega_{\mathrm{mod}}t)\end{pmatrix}
\end{equation}
with strength $B_0$ and (angular) rotation rate $\omega_{\mathrm{mod}}$. To isolate the performance of the synthesized magnetic field portion of the Hamiltonian in Eq.~\eqref{eqn:Hamiltonian}, the qubit is detuned from the boost cavity.

\subsection{Transverse Field (\texorpdfstring{\(B_x\)}{Bx})}\label{sec:Bx_cal}

We implement an effective $B_x$ by driving the qubit on resonance with a microwave drive through the readout cavity.
The Hamiltonian for a two level system under resonant drive is given by $H_q = \frac{\Omega}{2}\sigma_x$ where $\Omega$ is the Rabi rate, which is linearly proportional to the amplitude of the drive. The applied drive leads to oscillations between $\ket{g}$ and $\ket{e}$ at a rate $\Omega$. Measuring the $\ket{e}$-state population [$P(e)$] as a function of time and drive detuning produces a Rabi chevron pattern as shown in Fig.~\ref{fig:bx_cal}(a). As the detuning between the microwave drive and qubit frequency increases, the oscillation frequency of $P(e)$ increases with a corresponding decrease in the oscillation amplitude as given by \cite{atomic_physics_Foot}
\begin{equation}
    P(e) = \frac{\Omega^2}{\Omega^2+\Delta^2}\cos(\sqrt{\Omega^2+\Delta^2}t).
\end{equation}

We calibrate the effective transverse field ($B_x$) as a function of generator power (legend) and modulation amplitude (\(x\)-axis) using qubit spectroscopy, as shown in Fig.~\ref{fig:bx_cal}(b).
For each combination, we perform a standard Rabi measurement to extract the Rabi rate. 
As expected, the measured Rabi rate scales as the square root of the applied power meaning that a 6 dB increase in drive power is required to double the Rabi rate. The maximum Rabi rate of 110~MHz is limited by the output power of our generator and the linewidth of the cavity used to drive the qubit, and not by degradation of the qubit response, despite an anharmonicity of $\alpha=240$~MHz.

The required time-dependence of the transverse field is generated by modulating the amplitude sinusoidally: $\Omega(t)=\Omega_0\cos(\omega_{\mathrm{mod}}t)$.
The modulation is performed using an arbitrary waveform generator (AWG) connected to the IQ modulation port of the qubit generator, providing full control over the waveform. Simple amplitude modulation is not sufficient as the \emph{sign} of the applied field must change during the modulation period. 
We verify the time-dependence of the applied $x$-field by acquiring a Rabi chevron measurement in the presence of modulation, shown Fig.~\ref{fig:bx_cal}(c). The amplitude modulation of the drive [shown in Fig.~\ref{fig:bx_cal}(d)] is visible as a time-dependent oscillation frequency in the qubit population. In particular, the zero crossings in the field amplitude are identifiable as locations where the qubit dynamics freeze (at $t=$ 0.25 and 0.75~$\mu$s). 

Using the method described above, we can generate both static (DC) and dynamic (AC) effective $x$-fields on the transmon with amplitudes reaching 110 MHz, significantly exceeding the native boost cavity-qubit coupling of 13 MHz, as required by Eq.~\eqref{eqn:topo_regime} when \(n\) is moderately large.


\subsection{Longitudinal Field (\texorpdfstring{\(B_z\)}{Bz})}\label{sec:Bz_cal}

\begin{figure}[h]
    \centering
    \includegraphics[width=\linewidth]{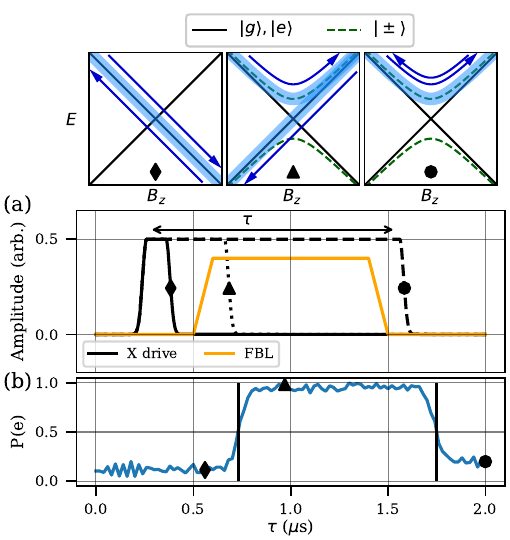}
    \caption{Calibration of the delay between $x$ and $z$ components of the applied field using Landau-Zener transitions.
    (a)~Pulse sequence used to determine the delay: a fixed 1~$\mu$s flux bias pulse (FBL, gold) is applied to the qubit to ramp it through resonance with a microwave drive (black) of variable duration $\tau$. Resonance occurs at the middle of the FBL pulse ramp such that both the starting and ending qubit frequency are far off resonance with the applied drive. Insets: energy level diagrams for the bare states ($\ket{g},\ket{e}$, black) and coupled states ($\ket{\pm}$ dashed green) of Eq.~\eqref{eq:lz}. Blue highlights (arrows) show the qubit dynamics in 3 cases: first, the microwave drive does not overlap with the FBL pulse (diamond) producing no LZ transition; second, the microwave drive overlaps only with the leading edge of the FBL pulse leading to one LZ transition (triangle); third, the microwave drive overlaps with both the leading and trailing edges of the FBL pulse leading to two LZ transitions (circle).
    (b) Measured qubit population as $\tau$ is varied. 
    The line delay is extracted by comparing the measured time interval of the qubit excitation to the position of the FBL pulse in the schedule.
    }
    \label{fig:bz_cal}
\end{figure}

The $z$-component of the rotating field is implemented by modulating the qubit frequency about a DC set point $\omega_{q_0}$.
An external magnet with large dynamic range and low bandwidth is used to set $\omega_{q_0}$, while an on-chip FBL (driven by the AWG) applies a sinusoidal frequency modulation to the qubit frequency, effectively generating
\begin{equation}
    B_z(t) = B_0 \sin(\omega_{\mathrm{mod}}t),
\end{equation}
where $B_0$ and $\omega_{\mathrm{mod}}$ are the field strength and rotation rate respectively.

Just as with the $B_x$ calibration, we first calibrate the size of the frequency shift. This is performed using qubit spectroscopy during a long (1~$\mu$s) flux pulse as a function of pulse amplitude from the AWG.

The $x$ and $z$ components of the applied field originate from two different control processes and natively have an uncontrolled delay between them.
However, they can be synchronized using two measurement protocols which convert the in-situ delay between the components into qubit-state population.
The first is a time-resolved spectroscopy measurement of the qubit to map out the flux bias pulse~\cite{cryoscope}, and the second is a measurement of Landau-Zener transitions~\cite{CohenTannoudji} which occur when the qubit frequency is adiabatically ramped through resonance with an applied drive. 
While the first method provides a full picture of the FBL pulse, it is time intensive, requiring a spectroscopic measurement at every time interval. We use the full time-dependent characterization to measure the filter induced distortions on the FBL pulse in Appendix~\ref{appendix:precompensation}. 
The Landau-Zener (LZ) based measurement, on the other hand, only requires a single frequency measurement making it significantly faster. 

LZ transitions occur when two coupled energy levels are swept through resonance~\cite{CohenTannoudji}.
Here we use an analogous process in the presence of AC drives, known in this context as adiabatic rapid passage, to convert the time-dependent qubit detuning into a qubit excitation which can be readily measured.
Consider the following coupled two-level system Hamiltonian: 
\begin{equation}\label{eq:lz}
    H = \frac{1}{2}\begin{pmatrix}
        B_z & B_x \\ B_x &-B_z        
    \end{pmatrix},
\end{equation}
where $B_x$ is the driving strength and $B_z$ is the detuning between the qubit and the drive. The eigenstates, $\ket{\pm}$, of this Hamiltonian are admixtures of the bare states $\ket{g},\ket{e}$ with a detuning dependent mixing fraction. As $B_z$ is swept through zero, the character of the ground state ($\ket{-}$) changes from $\ket{g}$ to $\ket{e}$. By slowly sweeping the detuning between the levels ($\dot B_z\ll B_x^2$) it is possible to perform a high-fidelity population transfer from one level to the other without requiring a $\pi$-pulse. The full energy diagram for the LZ transition is shown in the insets of Fig.~\ref{fig:bz_cal}, where the bare states are shown as the solid black lines and the coupled states are indicated with the dashed green lines.

Our measurement of the line delay utilizes LZ transitions to locate the times at which the qubit crosses resonance with an applied drive.
In particular, we apply a 1~$\mu$s trapezoidal flux bias pulse [shown in gold in Fig.~\ref{fig:bz_cal}(a)] of known amplitude and apply an off-resonant drive on the qubit [shown in black in Fig.~\ref{fig:bz_cal}(a)]. 
Schematically, we can write the pulses as:
\begin{align}
    B_x(t) & = \Omega~ b_{[0,\tau]}(t), \\
    B_z(t) & = \Delta[-1+2b_{[\tau_{\mathrm{del}},\tau_{\mathrm{del}}+1\,\mu\mathrm{s}]}(t)],
\end{align}
where $\Omega, \Delta$ are the drive strength and qubit detuning respectively, 
$b_{[\tau_1,\tau_2]}(t)$ is a unit amplitude step function (with linear on/off ramps) supported in the time interval \([\tau_1,\tau_2]\), and $\tau_{\mathrm{del}}>0$ is the unknown delay between the lines. An LZ transition occurs when $B_z$ changes sign and $B_x$ is non-zero. The delay time $\tau_{\mathrm{del}}$ can then be found by measuring the qubit population as a function of $\tau$ and mapping out the interval $[\tau_{\mathrm{del}},\tau_{\mathrm{del}}+1\,\mu\mathrm{s}]$ as shown in Fig.~\ref{fig:bz_cal}(b). 

We consider three qualitatively different regimes for the microwave pulse duration corresponding to zero, one, and two LZ transitions respectively in the insets of Fig.~\ref{fig:bz_cal}.
At short hold times ($\tau<\tau_{\mathrm{del}}$, diamond marker), the microwave drive and FBL pulse are temporally separated, leading the qubit to ramp through resonance with no applied drive and stay in the ground state at all times. At intermediate times ($\tau\in[\tau_\mathrm{del},\tau_{\mathrm{del}}+1\mu\mathrm{s}]$, triangle marker), the qubit ramps through resonance with the applied drive on during the rising edge of the FBL pulse, leading to an LZ transition to the $\ket{e}$ state. The falling edge of the FBL pulse occurs with no applied drive, leaving the qubit in $\ket{e}$. Finally, at long hold times ($\tau>\tau_{\mathrm{del}}+1\mu\mathrm{s}$, circle marker), both the rising and falling edges of the FBL pulse occur with an applied microwave drive, leading to two LZ transitions which bring the state back to $\ket{g}$ at the end of the measurement.

The methods described above provide a reliable way to synthesize $z$-fields as well as set the phase relationship between the $x$ and $z$-components of the magnetic field. For the circularly polarized field described in the introduction, the two components are driven with a 90-degree phase shift, although in principle any polarization is achievable with this toolkit. A phase coherent $y$-drive can be generated by using the Q-quadrature of the microwave signal. 

\section{Transmons in strong synthetic fields}\label{sec:following}

\begin{figure}[t]
	\begin{center}
		\includegraphics[width=\linewidth]{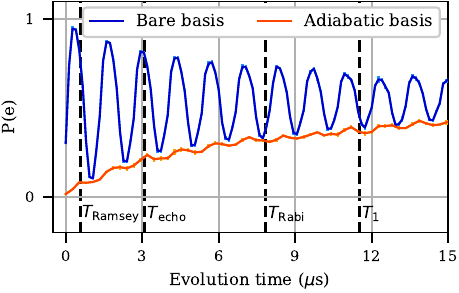}
	\end{center}
	\vspace{-0.6cm}
	\caption{\label{fig:time_scale_comp} 
    Measurement of the qubit adiabatically following the rotating field in the bare qubit basis (blue trace) and in the basis of the rotating field (orange).
    In both measurements, the field following time ($T_d=12$ $\mu$s) exceeds the other coherence times of the qubit ($T_{\mathrm{Ramsey}},T_{\mathrm{echo}},T_{\mathrm{Rabi}}$) and is comparable the qubit lifetime $T_1$ (11.5~$\mu$s). The qubit coherence times are indicated by the dashed lines, measurement data is shown in Appendix \ref{appendix:qubit_timescales}. The bare qubit basis measurements are acquired by an approximate diabatic shutoff of the fields and the adiabatic basis measurements are acquired by mapping the field $\ket{\pm}$ states back onto $\ket{g},\ket{e}$ \cite{dissipation_paper}. 
    } 
\end{figure}
\begin{figure*}
    \centering
    \includegraphics[width=\linewidth]{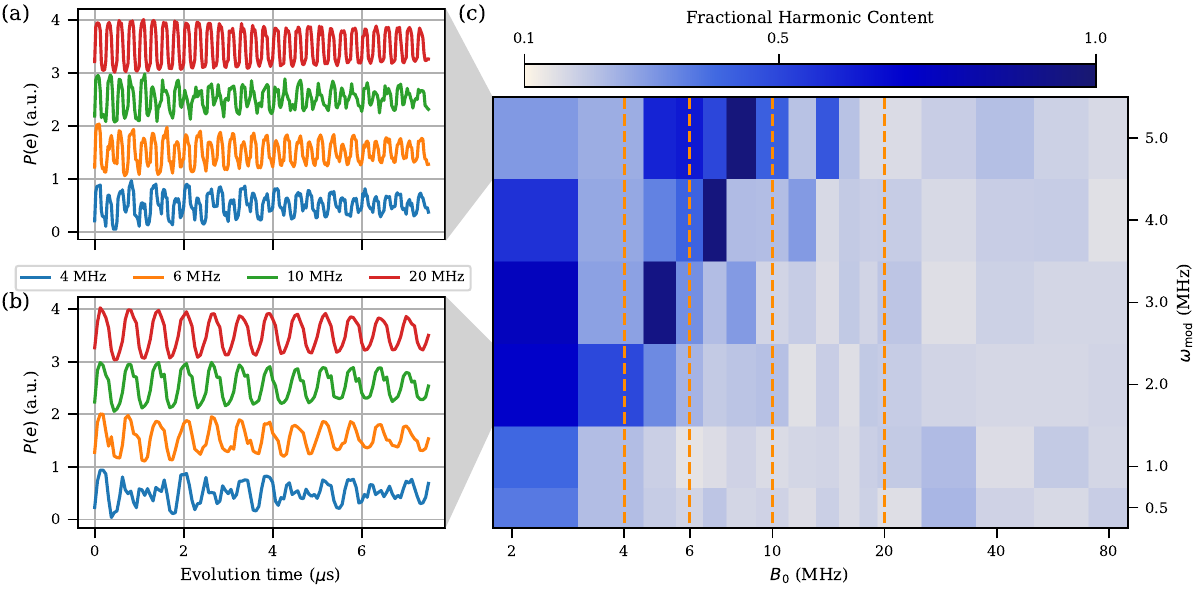}
    \caption{Degree of adiabatic following of external field. (a) Bare-qubit basis measurement versus time with a fast rotation rate $\omega_{\mathrm{mod}}=5$~MHz and $B_0=$ 4, 6, 10, 20~MHz [shown in the legend of (a), (b)]. At large $B_0$ (20 MHz), the qubit state is effectively pinned to the rotating field as demonstrated by the sinusoidal oscillations with rate $\omega_{\mathrm{mod}}$. As $B_0$ decreases, the qubit no longer adiabatically follows the rotating field; instead, it rotates with the field while nutating about it, creating multiple frequency components in the traces. (b) Bare qubit basis traces for a slower rotation rate $\omega_{\mathrm{mod}}=2$~MHz, showing onset of good following around 6 MHz as opposed to the 20 MHz field required for the faster rotation rate in panel (a). (c) Fractional harmonic content [as defined in Eq.~\eqref{eqn:metric}] as a function of applied field strength and rotation rate measured in the adiabatic basis. Lower values indicate smaller deviations from adiabatic following. For $B_0\gtrsim 2.5~ \omega_{\mathrm{mod}}$, the system follows the applied field with no observed dependence on the field size or rotation rate.
    }
    \label{fig:adiabatic_sweep}
\end{figure*}

We confirm the simultaneous operation of the synchronized effective $B$-field components and adiabatic following of the qubit by measuring the $\sigma_x$ and $\sigma_z$ components as a function of evolution time [Fig.~\ref{fig:pumping_schematic}(b)].
The measurement of the qubit state is acquired by evolving the qubit for a variable time under a field of strength $B_0$ rotating at rate $\omega_{\mathrm{mod}}$ before shutting off the fields diabatically and measuring the qubit state. The qubit is initialized in an eigenstate with $+1$ projection along the field before the field rotation begins to reduce nutation about the field (see Appendix \ref{appendix:initialization}). The $x$-projection is measured by applying a $\pi/2$-pulse after the field shut-off to rotate the measurement axis. In this measurement scheme, the $x$ and $z$-components of the spin oscillate at frequency $\omega_{\mathrm{mod}}$ with a 90-degree phase shift as expected from a spin with $+1$ projection along $\vec{B}(t)$ and rotating in the $xz$ plane. 

In addition to the measurement described above, we perform complementary measurements in the adiabatic basis of the rotating field (i.e. the instantaneous eigenbasis of $H$) which directly measure the degree of adiabatic following as a function of time.
This measurement is performed by adiabatically rotating the final field back to a fixed point on the Bloch sphere (in our case back to $-z$) before performing readout \cite{dissipation_paper}. For perfect readout and decoherence-free adiabatic evolution, the qubit stays in the eigenstate with $+1$ projection along the field and is mapped back to the $\ket{g}$ state for measurement. A comparison between the two measurement methods is shown in Fig.~\ref{fig:time_scale_comp}.
Both measurements evolve towards a mixed state at long evolution times, with the adiabatic basis measurement tracing out the envelope of the bare-qubit basis oscillatory data.

Measurements in either basis (bare-qubit basis or adiabatic basis) demonstrate that the field following time ($T_d$) far exceeds the qubit phase coherence times ($T_{2},T_{\mathrm{echo}}$), despite the flux sensitive nature of our architecture, and that $T_d$ is comparable to the depolarization time, $T_1$, as shown in Fig.~\ref{fig:time_scale_comp}.
We also measure an improvement compared to the Rabi time, which is less sensitive to the high-frequency noise on the qubit transition frequency but is limited by slow frequency qubit drifts \cite{ithier2005decoherence}. The long field following time indicates that the adiabatic following is effectively insensitive to $z$-field noise on the qubit transition, reminiscent of the spin-locking time in NMR-systems \cite{slichter_NMR}.

The timescale over which the qubit successfully adiabatically follows the drive is a critical parameter for performance of the topological energy pump. In practice, this can be limited by two qualitatively different types of effects. First, coupling of the qubit to the environment will eventually lead to incoherent mixing of the two adiabatic eigenstates, no matter how slowly the Hamiltonian is varied. 
Second, changing the Hamiltonian too quickly leads to partially diabatic transitions in the qubit. In this case, the qubit undergoes coherent and reproducible nutation about the applied field. 
We refer to these coherent effects as breakdown of adiabaticity since these are a property of the drive Hamiltonian and distinct from dissipation-induced effects.

Below, we measure the field following over a range of field amplitudes (2-80 MHz) and field rotation rates (0.5-5 MHz) in both the bare and adiabatic bases, and show that for large enough fields, the qubit follows the applied field, without breakdown of adiabaticity, independent of the rotation rate and field amplitude.

Ideal, dissipation-limited, adiabatic following is identifiable in the bare-basis measurements as a decaying oscillation with frequency $\omega_{\mathrm{mod}}$, as seen in the large field traces in Fig.~\ref{fig:adiabatic_sweep}(a-b). For the adiabatic basis measurements, ideal following is characterized by an exponential response. On the other hand, breakdown in adiabatic following at low $B_0$ or high $\omega_{\mathrm{mod}}$ is marked by the appearance of additional oscillations in the qubit population as the qubit nutates about the applied field.

As a result, the Fourier spectrum of the adiabatic basis measurements provides a simple measurement of the degree of adiabaticity.
We quantify this by the fraction of the spectrum that lies above a cutoff frequency corresponding to the width of the ideal Lorentzian response centered at zero frequency. 
This metric, termed fractional harmonic content, is defined as,
\begin{equation}\label{eqn:metric}
    F = \frac{\int_{\omega_c}^{\omega_s/2} \mathrm{d} \omega|f(\omega)|^2}{\int_{0}^{\omega_s/2} \mathrm{d}\omega |f(\omega)|^2}, 
\end{equation}
where $\omega_c$ is the cutoff frequency ($2\pi\times 167$~kHz), $\omega_s$ is the sampling rate of the data, and $|f(\omega)|^2$ is the power spectrum of the data. For ideal following, this metric approaches zero whereas a breakdown in adiabatic following corresponds to a value close to one. We compute this metric over a range of field amplitudes and rotation rates, in Fig.~\ref{fig:adiabatic_sweep}(c). Adiabatic following appears to break down for $B_0\lesssim2.5~\omega_{\mathrm{mod}}$.

\section{Conclusion}

In this article, we have demonstrated an experimental toolkit to synthesize large rotating effective magnetic fields for a transmon qubit. The effective field is synthesized by a combination of amplitude modulation of a resonant microwave drive ($x$-component) and sinusoidal frequency modulation of the qubit ($z$-component). We achieved time-dependent fields with peak amplitude exceeding 100 MHz, far exceeding the qubit-boost cavity coupling $g$ of 13 MHz. We also provide a simple method to combine the field components into a controlled rotating field. 

We show that, for large fields, the qubit adiabatically follows the applied field by performing measurements both in the bare qubit basis (where the population oscillates at $\omega_{\mathrm{mod}}$) and in the adiabatic basis (where the instantaneous eigenstates of the applied field are mapped onto $\ket{g},\ket{e}$). The field following time ($T_d$) far exceeds the qubit phase coherence times and Rabi time indicating a relative insensitivity to wide band frequency noise. 

Finally, we map out the adiabatic following regime versus $B_0$ (the field amplitude) and $\omega_{\mathrm{mod}}$ (the field rotation rate) and show that the adiabatic following is insensitive to $B_0$ and $\omega_{\mathrm{mod}}$ when $B_0\gtrsim 2.5~\omega_{\mathrm{mod}}$. 

The tools developed in this article can be used to drive the qubit in the desired regime for the topological pump proposed in \cite{Long2022b}. In addition, we show that the flux-tunable architecture does not severely limit the performance of the coherent drives as the large field amplitudes reduce the sensitivity of the qubit to flux noise. This work provides a clear pathway towards experimentally realizing topological energy pumps and direct observation of the quantized energy current. 

Recent work in Ref.~\cite{dissipation_paper} developed a protocol to further reduce the sensitivity of the pump to $T_1$ decoherence using a secondary lossy cavity to stabilize the pumping state. With the exception of the additional cavity, the proposed protocol of Ref.~\cite{dissipation_paper} relies on the same configuration of driving fields presented here allowing for another method to experimentally observe the topological energy pump using the experimental toolkit developed here. 

\begin{acknowledgments}
We thank Ben Cochran for contributions to device modeling; IBK Adisa for assistance with early measurements; and Zhiyin Tu and Gabriel Chiselenko for assistance with figure preparation. AC thanks the Max Planck Institute for the Physics of Complex Systems for its hospitality.
This work was supported by the ARL (Grant no. W911NF-19-2-0181 and W911NF-17-S-0003), the AFOSR (Grant no. FA9550-24-1-0121), and the NSF (QLCI grant OMA-2120757). QY received support from AFOSR grant no. FA9550-21-1-0129. MR received support from the LPS graduate fellowship, LPS Grant no. H9823022C0069, and ARCS.
DL was supported by the Air Force (AFOSR grant no. FA9550-20-1-0235), the Laboratory for Physical Sciences (through their support of the Condensed Matter Theory Center at the University of Maryland),
a Stanford QFARM Bloch postdoctoral fellowship, and the Packard Foundation through a Packard Fellowship in Science and Engineering (PI: Vedika Khemani).


\end{acknowledgments}

\bibliographystyle{apsrev4-2}
\bibliography{refs.bib}

\begin{thebibliography}{57}%
\makeatletter
\providecommand \@ifxundefined [1]{%
 \@ifx{#1\undefined}
}%
\providecommand \@ifnum [1]{%
 \ifnum #1\expandafter \@firstoftwo
 \else \expandafter \@secondoftwo
 \fi
}%
\providecommand \@ifx [1]{%
 \ifx #1\expandafter \@firstoftwo
 \else \expandafter \@secondoftwo
 \fi
}%
\providecommand \natexlab [1]{#1}%
\providecommand \enquote  [1]{``#1''}%
\providecommand \bibnamefont  [1]{#1}%
\providecommand \bibfnamefont [1]{#1}%
\providecommand \citenamefont [1]{#1}%
\providecommand \href@noop [0]{\@secondoftwo}%
\providecommand \href [0]{\begingroup \@sanitize@url \@href}%
\providecommand \@href[1]{\@@startlink{#1}\@@href}%
\providecommand \@@href[1]{\endgroup#1\@@endlink}%
\providecommand \@sanitize@url [0]{\catcode `\\12\catcode `\$12\catcode `\&12\catcode `\#12\catcode `\^12\catcode `\_12\catcode `\%12\relax}%
\providecommand \@@startlink[1]{}%
\providecommand \@@endlink[0]{}%
\providecommand \url  [0]{\begingroup\@sanitize@url \@url }%
\providecommand \@url [1]{\endgroup\@href {#1}{\urlprefix }}%
\providecommand \urlprefix  [0]{URL }%
\providecommand \Eprint [0]{\href }%
\providecommand \doibase [0]{http://dx.doi.org/}%
\providecommand \selectlanguage [0]{\@gobble}%
\providecommand \bibinfo  [0]{\@secondoftwo}%
\providecommand \bibfield  [0]{\@secondoftwo}%
\providecommand \translation [1]{[#1]}%
\providecommand \BibitemOpen [0]{}%
\providecommand \bibitemStop [0]{}%
\providecommand \bibitemNoStop [0]{.\EOS\space}%
\providecommand \EOS [0]{\spacefactor3000\relax}%
\providecommand \BibitemShut  [1]{\csname bibitem#1\endcsname}%
\let\auto@bib@innerbib\@empty
\bibitem [{\citenamefont {Long}\ \emph {et~al.}(2022)\citenamefont {Long}, \citenamefont {Crowley}, \citenamefont {Koll\'ar},\ and\ \citenamefont {Chandran}}]{Long2022b}%
  \BibitemOpen
  \bibfield  {author} {\bibinfo {author} {\bibfnamefont {D.~M.}\ \bibnamefont {Long}}, \bibinfo {author} {\bibfnamefont {P.~J.~D.}\ \bibnamefont {Crowley}}, \bibinfo {author} {\bibfnamefont {A.~J.}\ \bibnamefont {Koll\'ar}}, \ and\ \bibinfo {author} {\bibfnamefont {A.}~\bibnamefont {Chandran}},\ }\href {\doibase 10.1103/PhysRevLett.128.183602} {\bibfield  {journal} {\bibinfo  {journal} {Phys. Rev. Lett.}\ }\textbf {\bibinfo {volume} {128}},\ \bibinfo {pages} {183602} (\bibinfo {year} {2022})}\BibitemShut {NoStop}%
\bibitem [{\citenamefont {Born}\ and\ \citenamefont {Fock}(1928)}]{Born1928}%
  \BibitemOpen
  \bibfield  {author} {\bibinfo {author} {\bibfnamefont {M.}~\bibnamefont {Born}}\ and\ \bibinfo {author} {\bibfnamefont {V.}~\bibnamefont {Fock}},\ }\href {\doibase 10.1007/BF01343193} {\bibfield  {journal} {\bibinfo  {journal} {Zeitschrift f{\"u}r Physik}\ }\textbf {\bibinfo {volume} {51}},\ \bibinfo {pages} {165} (\bibinfo {year} {1928})}\BibitemShut {NoStop}%
\bibitem [{\citenamefont {Comparat}(2009)}]{general_adiabatic_review}%
  \BibitemOpen
  \bibfield  {author} {\bibinfo {author} {\bibfnamefont {D.}~\bibnamefont {Comparat}},\ }\href {\doibase 10.1103/PhysRevA.80.012106} {\bibfield  {journal} {\bibinfo  {journal} {Phys. Rev. A}\ }\textbf {\bibinfo {volume} {80}},\ \bibinfo {pages} {012106} (\bibinfo {year} {2009})}\BibitemShut {NoStop}%
\bibitem [{\citenamefont {Farhi}\ \emph {et~al.}(2001)\citenamefont {Farhi}, \citenamefont {Goldstone}, \citenamefont {Gutmann}, \citenamefont {Lapan}, \citenamefont {Lundgren},\ and\ \citenamefont {Preda}}]{adiabatic_QC_Farhi_2001}%
  \BibitemOpen
  \bibfield  {author} {\bibinfo {author} {\bibfnamefont {E.}~\bibnamefont {Farhi}}, \bibinfo {author} {\bibfnamefont {J.}~\bibnamefont {Goldstone}}, \bibinfo {author} {\bibfnamefont {S.}~\bibnamefont {Gutmann}}, \bibinfo {author} {\bibfnamefont {J.}~\bibnamefont {Lapan}}, \bibinfo {author} {\bibfnamefont {A.}~\bibnamefont {Lundgren}}, \ and\ \bibinfo {author} {\bibfnamefont {D.}~\bibnamefont {Preda}},\ }\href {\doibase 10.1126/science.1057726} {\bibfield  {journal} {\bibinfo  {journal} {Science}\ }\textbf {\bibinfo {volume} {292}},\ \bibinfo {pages} {472} (\bibinfo {year} {2001})},\ \Eprint {http://arxiv.org/abs/https://www.science.org/doi/pdf/10.1126/science.1057726}{https://www.science.org/doi/pdf/10.1126/science.1057726}\BibitemShut {NoStop}%
\bibitem [{\citenamefont {Kolodrubetz}\ \emph {et~al.}(2017)\citenamefont {Kolodrubetz}, \citenamefont {Sels}, \citenamefont {Mehta},\ and\ \citenamefont {Polkovnikov}}]{adiabatic_geometry_review_Kolodrubetz_2017}%
  \BibitemOpen
  \bibfield  {author} {\bibinfo {author} {\bibfnamefont {M.}~\bibnamefont {Kolodrubetz}}, \bibinfo {author} {\bibfnamefont {D.}~\bibnamefont {Sels}}, \bibinfo {author} {\bibfnamefont {P.}~\bibnamefont {Mehta}}, \ and\ \bibinfo {author} {\bibfnamefont {A.}~\bibnamefont {Polkovnikov}},\ }\href {\doibase https://doi.org/10.1016/j.physrep.2017.07.001} {\bibfield  {journal} {\bibinfo  {journal} {Physics Reports}\ }\textbf {\bibinfo {volume} {697}},\ \bibinfo {pages} {1} (\bibinfo {year} {2017})},\ \bibinfo {note} {geometry and non-adiabatic response in quantum and classical systems}\BibitemShut {NoStop}%
\bibitem [{\citenamefont {Albash}\ and\ \citenamefont {Lidar}(2018)}]{adiabaticQC_review_Albash_2018}%
  \BibitemOpen
  \bibfield  {author} {\bibinfo {author} {\bibfnamefont {T.}~\bibnamefont {Albash}}\ and\ \bibinfo {author} {\bibfnamefont {D.~A.}\ \bibnamefont {Lidar}},\ }\href {\doibase 10.1103/RevModPhys.90.015002} {\bibfield  {journal} {\bibinfo  {journal} {Rev. Mod. Phys.}\ }\textbf {\bibinfo {volume} {90}},\ \bibinfo {pages} {015002} (\bibinfo {year} {2018})}\BibitemShut {NoStop}%
\bibitem [{\citenamefont {Weinberg}\ \emph {et~al.}(2017)\citenamefont {Weinberg}, \citenamefont {Bukov}, \citenamefont {D’Alessio}, \citenamefont {Polkovnikov}, \citenamefont {Vajna},\ and\ \citenamefont {Kolodrubetz}}]{adiabatic_review_periodically_drive_Weinberg_2017}%
  \BibitemOpen
  \bibfield  {author} {\bibinfo {author} {\bibfnamefont {P.}~\bibnamefont {Weinberg}}, \bibinfo {author} {\bibfnamefont {M.}~\bibnamefont {Bukov}}, \bibinfo {author} {\bibfnamefont {L.}~\bibnamefont {D’Alessio}}, \bibinfo {author} {\bibfnamefont {A.}~\bibnamefont {Polkovnikov}}, \bibinfo {author} {\bibfnamefont {S.}~\bibnamefont {Vajna}}, \ and\ \bibinfo {author} {\bibfnamefont {M.}~\bibnamefont {Kolodrubetz}},\ }\href {\doibase https://doi.org/10.1016/j.physrep.2017.05.003} {\bibfield  {journal} {\bibinfo  {journal} {Physics Reports}\ }\textbf {\bibinfo {volume} {688}},\ \bibinfo {pages} {1} (\bibinfo {year} {2017})}\BibitemShut {NoStop}%
\bibitem [{\citenamefont {Thouless}(1983)}]{Thouless1983pump}%
  \BibitemOpen
  \bibfield  {author} {\bibinfo {author} {\bibfnamefont {D.~J.}\ \bibnamefont {Thouless}},\ }\href {\doibase 10.1103/PhysRevB.27.6083} {\bibfield  {journal} {\bibinfo  {journal} {Phys. Rev. B}\ }\textbf {\bibinfo {volume} {27}},\ \bibinfo {pages} {6083} (\bibinfo {year} {1983})}\BibitemShut {NoStop}%
\bibitem [{\citenamefont {Citro}\ and\ \citenamefont {Aidelsburger}(2023{\natexlab{a}})}]{Citro:2023_thoulesspumpreview}%
  \BibitemOpen
  \bibfield  {author} {\bibinfo {author} {\bibfnamefont {R.}~\bibnamefont {Citro}}\ and\ \bibinfo {author} {\bibfnamefont {M.}~\bibnamefont {Aidelsburger}},\ }\href {\doibase 10.1038/s42254-022-00545-0} {\bibfield  {journal} {\bibinfo  {journal} {Nature Reviews Physics}\ }\textbf {\bibinfo {volume} {5}},\ \bibinfo {pages} {87} (\bibinfo {year} {2023}{\natexlab{a}})}\BibitemShut {NoStop}%
\bibitem [{\citenamefont {Martin}\ \emph {et~al.}(2017)\citenamefont {Martin}, \citenamefont {Refael},\ and\ \citenamefont {Halperin}}]{Martin:2017aa}%
  \BibitemOpen
  \bibfield  {author} {\bibinfo {author} {\bibfnamefont {I.}~\bibnamefont {Martin}}, \bibinfo {author} {\bibfnamefont {G.}~\bibnamefont {Refael}}, \ and\ \bibinfo {author} {\bibfnamefont {B.}~\bibnamefont {Halperin}},\ }\href {\doibase 10.1103/PhysRevX.7.041008} {\bibfield  {journal} {\bibinfo  {journal} {Phys. Rev. X}\ }\textbf {\bibinfo {volume} {7}},\ \bibinfo {pages} {041008} (\bibinfo {year} {2017})}\BibitemShut {NoStop}%
\bibitem [{\citenamefont {Crowley}\ \emph {et~al.}(2019)\citenamefont {Crowley}, \citenamefont {Martin},\ and\ \citenamefont {Chandran}}]{Crowley:2019_classification}%
  \BibitemOpen
  \bibfield  {author} {\bibinfo {author} {\bibfnamefont {P.~J.~D.}\ \bibnamefont {Crowley}}, \bibinfo {author} {\bibfnamefont {I.}~\bibnamefont {Martin}}, \ and\ \bibinfo {author} {\bibfnamefont {A.}~\bibnamefont {Chandran}},\ }\href {\doibase 10.1103/PhysRevB.99.064306} {\bibfield  {journal} {\bibinfo  {journal} {Phys. Rev. B}\ }\textbf {\bibinfo {volume} {99}},\ \bibinfo {pages} {064306} (\bibinfo {year} {2019})}\BibitemShut {NoStop}%
\bibitem [{\citenamefont {Nathan}\ \emph {et~al.}(2019)\citenamefont {Nathan}, \citenamefont {Martin},\ and\ \citenamefont {Refael}}]{Nathan:2019_drivendiss}%
  \BibitemOpen
  \bibfield  {author} {\bibinfo {author} {\bibfnamefont {F.}~\bibnamefont {Nathan}}, \bibinfo {author} {\bibfnamefont {I.}~\bibnamefont {Martin}}, \ and\ \bibinfo {author} {\bibfnamefont {G.}~\bibnamefont {Refael}},\ }\href {\doibase 10.1103/PhysRevB.99.094311} {\bibfield  {journal} {\bibinfo  {journal} {Phys. Rev. B}\ }\textbf {\bibinfo {volume} {99}},\ \bibinfo {pages} {094311} (\bibinfo {year} {2019})}\BibitemShut {NoStop}%
\bibitem [{\citenamefont {Psaroudaki}\ and\ \citenamefont {Refael}(2021)}]{PSAROUDAKI2021}%
  \BibitemOpen
  \bibfield  {author} {\bibinfo {author} {\bibfnamefont {C.}~\bibnamefont {Psaroudaki}}\ and\ \bibinfo {author} {\bibfnamefont {G.}~\bibnamefont {Refael}},\ }\href {\doibase https://doi.org/10.1016/j.aop.2021.168553} {\bibfield  {journal} {\bibinfo  {journal} {Annals of Physics}\ }\textbf {\bibinfo {volume} {435}},\ \bibinfo {pages} {168553} (\bibinfo {year} {2021})},\ \bibinfo {note} {special issue on Philip W. Anderson}\BibitemShut {NoStop}%
\bibitem [{\citenamefont {Long}\ \emph {et~al.}(2021)\citenamefont {Long}, \citenamefont {Crowley},\ and\ \citenamefont {Chandran}}]{Long2021:class}%
  \BibitemOpen
  \bibfield  {author} {\bibinfo {author} {\bibfnamefont {D.~M.}\ \bibnamefont {Long}}, \bibinfo {author} {\bibfnamefont {P.~J.~D.}\ \bibnamefont {Crowley}}, \ and\ \bibinfo {author} {\bibfnamefont {A.}~\bibnamefont {Chandran}},\ }\href {\doibase 10.1103/PhysRevLett.126.106805} {\bibfield  {journal} {\bibinfo  {journal} {Phys. Rev. Lett.}\ }\textbf {\bibinfo {volume} {126}},\ \bibinfo {pages} {106805} (\bibinfo {year} {2021})}\BibitemShut {NoStop}%
\bibitem [{\citenamefont {Nathan}\ \emph {et~al.}(2021)\citenamefont {Nathan}, \citenamefont {Ge}, \citenamefont {Gazit}, \citenamefont {Rudner},\ and\ \citenamefont {Kolodrubetz}}]{Nathan2020b}%
  \BibitemOpen
  \bibfield  {author} {\bibinfo {author} {\bibfnamefont {F.}~\bibnamefont {Nathan}}, \bibinfo {author} {\bibfnamefont {R.}~\bibnamefont {Ge}}, \bibinfo {author} {\bibfnamefont {S.}~\bibnamefont {Gazit}}, \bibinfo {author} {\bibfnamefont {M.}~\bibnamefont {Rudner}}, \ and\ \bibinfo {author} {\bibfnamefont {M.}~\bibnamefont {Kolodrubetz}},\ }\href {\doibase 10.1103/PhysRevLett.127.166804} {\bibfield  {journal} {\bibinfo  {journal} {Phys. Rev. Lett.}\ }\textbf {\bibinfo {volume} {127}},\ \bibinfo {pages} {166804} (\bibinfo {year} {2021})}\BibitemShut {NoStop}%
\bibitem [{\citenamefont {Schroer}\ \emph {et~al.}(2014)\citenamefont {Schroer}, \citenamefont {Kolodrubetz}, \citenamefont {Kindel}, \citenamefont {Sandberg}, \citenamefont {Gao}, \citenamefont {Vissers}, \citenamefont {Pappas}, \citenamefont {Polkovnikov},\ and\ \citenamefont {Lehnert}}]{schroer_2014_topo_transition_spin_12}%
  \BibitemOpen
  \bibfield  {author} {\bibinfo {author} {\bibfnamefont {M.~D.}\ \bibnamefont {Schroer}}, \bibinfo {author} {\bibfnamefont {M.~H.}\ \bibnamefont {Kolodrubetz}}, \bibinfo {author} {\bibfnamefont {W.~F.}\ \bibnamefont {Kindel}}, \bibinfo {author} {\bibfnamefont {M.}~\bibnamefont {Sandberg}}, \bibinfo {author} {\bibfnamefont {J.}~\bibnamefont {Gao}}, \bibinfo {author} {\bibfnamefont {M.~R.}\ \bibnamefont {Vissers}}, \bibinfo {author} {\bibfnamefont {D.~P.}\ \bibnamefont {Pappas}}, \bibinfo {author} {\bibfnamefont {A.}~\bibnamefont {Polkovnikov}}, \ and\ \bibinfo {author} {\bibfnamefont {K.~W.}\ \bibnamefont {Lehnert}},\ }\href {\doibase 10.1103/PhysRevLett.113.050402} {\bibfield  {journal} {\bibinfo  {journal} {Phys. Rev. Lett.}\ }\textbf {\bibinfo {volume} {113}},\ \bibinfo {pages} {050402} (\bibinfo {year} {2014})}\BibitemShut {NoStop}%
\bibitem [{\citenamefont {Sridhar}\ \emph {et~al.}(2024)\citenamefont {Sridhar}, \citenamefont {Ghosh}, \citenamefont {Srinivasan}, \citenamefont {Miller},\ and\ \citenamefont {Dutt}}]{Sridhar2024photonic}%
  \BibitemOpen
  \bibfield  {author} {\bibinfo {author} {\bibfnamefont {S.~K.}\ \bibnamefont {Sridhar}}, \bibinfo {author} {\bibfnamefont {S.}~\bibnamefont {Ghosh}}, \bibinfo {author} {\bibfnamefont {D.}~\bibnamefont {Srinivasan}}, \bibinfo {author} {\bibfnamefont {A.~R.}\ \bibnamefont {Miller}}, \ and\ \bibinfo {author} {\bibfnamefont {A.}~\bibnamefont {Dutt}},\ }\href {\doibase 10.1038/s41567-024-02413-3} {\bibfield  {journal} {\bibinfo  {journal} {Nature Physics}\ }\textbf {\bibinfo {volume} {20}},\ \bibinfo {pages} {843} (\bibinfo {year} {2024})}\BibitemShut {NoStop}%
\bibitem [{\citenamefont {Koch}\ and\ \citenamefont {Budich}(2024)}]{Koch2024dissipative}%
  \BibitemOpen
  \bibfield  {author} {\bibinfo {author} {\bibfnamefont {F.}~\bibnamefont {Koch}}\ and\ \bibinfo {author} {\bibfnamefont {J.~C.}\ \bibnamefont {Budich}},\ }\href {\doibase 10.1103/PhysRevResearch.6.033124} {\bibfield  {journal} {\bibinfo  {journal} {Phys. Rev. Res.}\ }\textbf {\bibinfo {volume} {6}},\ \bibinfo {pages} {033124} (\bibinfo {year} {2024})}\BibitemShut {NoStop}%
\bibitem [{\citenamefont {Yuan}\ \emph {et~al.}(2018)\citenamefont {Yuan}, \citenamefont {Xiao}, \citenamefont {Lin},\ and\ \citenamefont {Fan}}]{Yuan2018ringresonator}%
  \BibitemOpen
  \bibfield  {author} {\bibinfo {author} {\bibfnamefont {L.}~\bibnamefont {Yuan}}, \bibinfo {author} {\bibfnamefont {M.}~\bibnamefont {Xiao}}, \bibinfo {author} {\bibfnamefont {Q.}~\bibnamefont {Lin}}, \ and\ \bibinfo {author} {\bibfnamefont {S.}~\bibnamefont {Fan}},\ }\href {\doibase 10.1103/PhysRevB.97.104105} {\bibfield  {journal} {\bibinfo  {journal} {Phys. Rev. B}\ }\textbf {\bibinfo {volume} {97}},\ \bibinfo {pages} {104105} (\bibinfo {year} {2018})}\BibitemShut {NoStop}%
\bibitem [{\citenamefont {Nathan}\ \emph {et~al.}(2022)\citenamefont {Nathan}, \citenamefont {Martin},\ and\ \citenamefont {Refael}}]{Nathan2022weyl}%
  \BibitemOpen
  \bibfield  {author} {\bibinfo {author} {\bibfnamefont {F.}~\bibnamefont {Nathan}}, \bibinfo {author} {\bibfnamefont {I.}~\bibnamefont {Martin}}, \ and\ \bibinfo {author} {\bibfnamefont {G.}~\bibnamefont {Refael}},\ }\href {\doibase 10.1103/PhysRevResearch.4.043060} {\bibfield  {journal} {\bibinfo  {journal} {Phys. Rev. Res.}\ }\textbf {\bibinfo {volume} {4}},\ \bibinfo {pages} {043060} (\bibinfo {year} {2022})}\BibitemShut {NoStop}%
\bibitem [{\citenamefont {Lantagne-Hurtubise}\ \emph {et~al.}(2024)\citenamefont {Lantagne-Hurtubise}, \citenamefont {Esin}, \citenamefont {Refael},\ and\ \citenamefont {Nathan}}]{Lantagne-Hurtubise2024graphene}%
  \BibitemOpen
  \bibfield  {author} {\bibinfo {author} {\bibfnamefont {E.}~\bibnamefont {Lantagne-Hurtubise}}, \bibinfo {author} {\bibfnamefont {I.}~\bibnamefont {Esin}}, \bibinfo {author} {\bibfnamefont {G.}~\bibnamefont {Refael}}, \ and\ \bibinfo {author} {\bibfnamefont {F.}~\bibnamefont {Nathan}},\ }\href {\doibase 10.1103/PhysRevB.110.L100305} {\bibfield  {journal} {\bibinfo  {journal} {Phys. Rev. B}\ }\textbf {\bibinfo {volume} {110}},\ \bibinfo {pages} {L100305} (\bibinfo {year} {2024})}\BibitemShut {NoStop}%
\bibitem [{\citenamefont {Luneau}\ \emph {et~al.}(2022)\citenamefont {Luneau}, \citenamefont {Dutreix}, \citenamefont {Ficheux}, \citenamefont {Delplace}, \citenamefont {Dou\ifmmode~\mbox{\c{c}}\else \c{c}\fi{}ot}, \citenamefont {Huard},\ and\ \citenamefont {Carpentier}}]{Luneau2022power}%
  \BibitemOpen
  \bibfield  {author} {\bibinfo {author} {\bibfnamefont {J.}~\bibnamefont {Luneau}}, \bibinfo {author} {\bibfnamefont {C.}~\bibnamefont {Dutreix}}, \bibinfo {author} {\bibfnamefont {Q.}~\bibnamefont {Ficheux}}, \bibinfo {author} {\bibfnamefont {P.}~\bibnamefont {Delplace}}, \bibinfo {author} {\bibfnamefont {B.}~\bibnamefont {Dou\ifmmode~\mbox{\c{c}}\else \c{c}\fi{}ot}}, \bibinfo {author} {\bibfnamefont {B.}~\bibnamefont {Huard}}, \ and\ \bibinfo {author} {\bibfnamefont {D.}~\bibnamefont {Carpentier}},\ }\href {\doibase 10.1103/PhysRevResearch.4.013169} {\bibfield  {journal} {\bibinfo  {journal} {Phys. Rev. Res.}\ }\textbf {\bibinfo {volume} {4}},\ \bibinfo {pages} {013169} (\bibinfo {year} {2022})}\BibitemShut {NoStop}%
\bibitem [{\citenamefont {Schwennicke}\ and\ \citenamefont {Yuen-Zhou}(2022)}]{Schwennicke2022enantioselective}%
  \BibitemOpen
  \bibfield  {author} {\bibinfo {author} {\bibfnamefont {K.}~\bibnamefont {Schwennicke}}\ and\ \bibinfo {author} {\bibfnamefont {J.}~\bibnamefont {Yuen-Zhou}},\ }\href {\doibase 10.1021/acs.jpclett.1c04161} {\bibfield  {journal} {\bibinfo  {journal} {The Journal of Physical Chemistry Letters}\ }\textbf {\bibinfo {volume} {13}},\ \bibinfo {pages} {2434} (\bibinfo {year} {2022})}\BibitemShut {NoStop}%
\bibitem [{\citenamefont {Cheng}\ \emph {et~al.}(2020)\citenamefont {Cheng}, \citenamefont {Prodan},\ and\ \citenamefont {Prodan}}]{Cheng_topo_pump_metamaterial_2020}%
  \BibitemOpen
  \bibfield  {author} {\bibinfo {author} {\bibfnamefont {W.}~\bibnamefont {Cheng}}, \bibinfo {author} {\bibfnamefont {E.}~\bibnamefont {Prodan}}, \ and\ \bibinfo {author} {\bibfnamefont {C.}~\bibnamefont {Prodan}},\ }\href {\doibase 10.1103/PhysRevLett.125.224301} {\bibfield  {journal} {\bibinfo  {journal} {Phys. Rev. Lett.}\ }\textbf {\bibinfo {volume} {125}},\ \bibinfo {pages} {224301} (\bibinfo {year} {2020})}\BibitemShut {NoStop}%
\bibitem [{\citenamefont {Kraus}\ \emph {et~al.}(2012)\citenamefont {Kraus}, \citenamefont {Lahini}, \citenamefont {Ringel}, \citenamefont {Verbin},\ and\ \citenamefont {Zilberberg}}]{quasi_crystal_photonic_Thouless}%
  \BibitemOpen
  \bibfield  {author} {\bibinfo {author} {\bibfnamefont {Y.~E.}\ \bibnamefont {Kraus}}, \bibinfo {author} {\bibfnamefont {Y.}~\bibnamefont {Lahini}}, \bibinfo {author} {\bibfnamefont {Z.}~\bibnamefont {Ringel}}, \bibinfo {author} {\bibfnamefont {M.}~\bibnamefont {Verbin}}, \ and\ \bibinfo {author} {\bibfnamefont {O.}~\bibnamefont {Zilberberg}},\ }\href {\doibase 10.1103/PhysRevLett.109.106402} {\bibfield  {journal} {\bibinfo  {journal} {Phys. Rev. Lett.}\ }\textbf {\bibinfo {volume} {109}},\ \bibinfo {pages} {106402} (\bibinfo {year} {2012})}\BibitemShut {NoStop}%
\bibitem [{\citenamefont {Nakajima}\ \emph {et~al.}(2016)\citenamefont {Nakajima}, \citenamefont {Tomita}, \citenamefont {Taie}, \citenamefont {Ichinose}, \citenamefont {Ozawa}, \citenamefont {Wang}, \citenamefont {Troyer},\ and\ \citenamefont {Takahashi}}]{fermion_Thouless_pump}%
  \BibitemOpen
  \bibfield  {author} {\bibinfo {author} {\bibfnamefont {S.}~\bibnamefont {Nakajima}}, \bibinfo {author} {\bibfnamefont {T.}~\bibnamefont {Tomita}}, \bibinfo {author} {\bibfnamefont {S.}~\bibnamefont {Taie}}, \bibinfo {author} {\bibfnamefont {T.}~\bibnamefont {Ichinose}}, \bibinfo {author} {\bibfnamefont {H.}~\bibnamefont {Ozawa}}, \bibinfo {author} {\bibfnamefont {L.}~\bibnamefont {Wang}}, \bibinfo {author} {\bibfnamefont {M.}~\bibnamefont {Troyer}}, \ and\ \bibinfo {author} {\bibfnamefont {Y.}~\bibnamefont {Takahashi}},\ }\href {\doibase 10.1038/nphys3622} {\bibfield  {journal} {\bibinfo  {journal} {Nature Physics}\ }\textbf {\bibinfo {volume} {12}},\ \bibinfo {pages} {296} (\bibinfo {year} {2016})}\BibitemShut {NoStop}%
\bibitem [{\citenamefont {Lohse}\ \emph {et~al.}(2018)\citenamefont {Lohse}, \citenamefont {Schweizer}, \citenamefont {Price}, \citenamefont {Zilberberg},\ and\ \citenamefont {Bloch}}]{Lohse2018}%
  \BibitemOpen
  \bibfield  {author} {\bibinfo {author} {\bibfnamefont {M.}~\bibnamefont {Lohse}}, \bibinfo {author} {\bibfnamefont {C.}~\bibnamefont {Schweizer}}, \bibinfo {author} {\bibfnamefont {H.~M.}\ \bibnamefont {Price}}, \bibinfo {author} {\bibfnamefont {O.}~\bibnamefont {Zilberberg}}, \ and\ \bibinfo {author} {\bibfnamefont {I.}~\bibnamefont {Bloch}},\ }\href {\doibase 10.1038/nature25000} {\bibfield  {journal} {\bibinfo  {journal} {Nature (London)}\ }\textbf {\bibinfo {volume} {553}},\ \bibinfo {pages} {55} (\bibinfo {year} {2018})}\BibitemShut {NoStop}%
\bibitem [{\citenamefont {Lohse}\ \emph {et~al.}(2016)\citenamefont {Lohse}, \citenamefont {Schweizer}, \citenamefont {Zilberberg}, \citenamefont {Aidelsburger},\ and\ \citenamefont {Bloch}}]{Thouless_pump_Lohse_2016}%
  \BibitemOpen
  \bibfield  {author} {\bibinfo {author} {\bibfnamefont {M.}~\bibnamefont {Lohse}}, \bibinfo {author} {\bibfnamefont {C.}~\bibnamefont {Schweizer}}, \bibinfo {author} {\bibfnamefont {O.}~\bibnamefont {Zilberberg}}, \bibinfo {author} {\bibfnamefont {M.}~\bibnamefont {Aidelsburger}}, \ and\ \bibinfo {author} {\bibfnamefont {I.}~\bibnamefont {Bloch}},\ }\href {\doibase 10.1038/nphys3584} {\bibfield  {journal} {\bibinfo  {journal} {Nature Physics}\ }\textbf {\bibinfo {volume} {12}},\ \bibinfo {pages} {350} (\bibinfo {year} {2016})}\BibitemShut {NoStop}%
\bibitem [{\citenamefont {Cerjan}\ \emph {et~al.}(2020)\citenamefont {Cerjan}, \citenamefont {Wang}, \citenamefont {Huang}, \citenamefont {Chen},\ and\ \citenamefont {Rechtsman}}]{Thouless_pump_disordered_photonics_Cerjan}%
  \BibitemOpen
  \bibfield  {author} {\bibinfo {author} {\bibfnamefont {A.}~\bibnamefont {Cerjan}}, \bibinfo {author} {\bibfnamefont {M.}~\bibnamefont {Wang}}, \bibinfo {author} {\bibfnamefont {S.}~\bibnamefont {Huang}}, \bibinfo {author} {\bibfnamefont {K.~P.}\ \bibnamefont {Chen}}, \ and\ \bibinfo {author} {\bibfnamefont {M.~C.}\ \bibnamefont {Rechtsman}},\ }\href {\doibase 10.1038/s41377-020-00408-2} {\bibfield  {journal} {\bibinfo  {journal} {Light: Science {\&} Applications}\ }\textbf {\bibinfo {volume} {9}},\ \bibinfo {pages} {178} (\bibinfo {year} {2020})}\BibitemShut {NoStop}%
\bibitem [{\citenamefont {J{\"u}rgensen}\ \emph {et~al.}(2021)\citenamefont {J{\"u}rgensen}, \citenamefont {Mukherjee},\ and\ \citenamefont {Rechtsman}}]{Jürgensen_nonlinear_Thouless_pump_2021}%
  \BibitemOpen
  \bibfield  {author} {\bibinfo {author} {\bibfnamefont {M.}~\bibnamefont {J{\"u}rgensen}}, \bibinfo {author} {\bibfnamefont {S.}~\bibnamefont {Mukherjee}}, \ and\ \bibinfo {author} {\bibfnamefont {M.~C.}\ \bibnamefont {Rechtsman}},\ }\href {\doibase 10.1038/s41586-021-03688-9} {\bibfield  {journal} {\bibinfo  {journal} {Nature}\ }\textbf {\bibinfo {volume} {596}},\ \bibinfo {pages} {63} (\bibinfo {year} {2021})}\BibitemShut {NoStop}%
\bibitem [{\citenamefont {Mostaan}\ \emph {et~al.}(2022)\citenamefont {Mostaan}, \citenamefont {Grusdt},\ and\ \citenamefont {Goldman}}]{Mostaan_nonlinear_Thouless_soliton_atoms_2022}%
  \BibitemOpen
  \bibfield  {author} {\bibinfo {author} {\bibfnamefont {N.}~\bibnamefont {Mostaan}}, \bibinfo {author} {\bibfnamefont {F.}~\bibnamefont {Grusdt}}, \ and\ \bibinfo {author} {\bibfnamefont {N.}~\bibnamefont {Goldman}},\ }\href {\doibase 10.1038/s41467-022-33478-4} {\bibfield  {journal} {\bibinfo  {journal} {Nature Communications}\ }\textbf {\bibinfo {volume} {13}},\ \bibinfo {pages} {5997} (\bibinfo {year} {2022})}\BibitemShut {NoStop}%
\bibitem [{\citenamefont {Grinberg}\ \emph {et~al.}(2020)\citenamefont {Grinberg}, \citenamefont {Lin}, \citenamefont {Harris}, \citenamefont {Benalcazar}, \citenamefont {Peterson}, \citenamefont {Hughes},\ and\ \citenamefont {Bahl}}]{grinberg_robust_2020}%
  \BibitemOpen
  \bibfield  {author} {\bibinfo {author} {\bibfnamefont {I.~H.}\ \bibnamefont {Grinberg}}, \bibinfo {author} {\bibfnamefont {M.}~\bibnamefont {Lin}}, \bibinfo {author} {\bibfnamefont {C.}~\bibnamefont {Harris}}, \bibinfo {author} {\bibfnamefont {W.~A.}\ \bibnamefont {Benalcazar}}, \bibinfo {author} {\bibfnamefont {C.~W.}\ \bibnamefont {Peterson}}, \bibinfo {author} {\bibfnamefont {T.~L.}\ \bibnamefont {Hughes}}, \ and\ \bibinfo {author} {\bibfnamefont {G.}~\bibnamefont {Bahl}},\ }\href {\doibase 10.1038/s41467-020-14804-0} {\bibfield  {journal} {\bibinfo  {journal} {Nature Communications}\ }\textbf {\bibinfo {volume} {11}},\ \bibinfo {pages} {974} (\bibinfo {year} {2020})}\BibitemShut {NoStop}%
\bibitem [{\citenamefont {Walter}\ \emph {et~al.}(2023)\citenamefont {Walter}, \citenamefont {Zhu}, \citenamefont {Gächter}, \citenamefont {Minguzzi}, \citenamefont {Roschinski}, \citenamefont {Sandholzer}, \citenamefont {Viebahn},\ and\ \citenamefont {Esslinger}}]{walter_quantization_2023}%
  \BibitemOpen
  \bibfield  {author} {\bibinfo {author} {\bibfnamefont {A.-S.}\ \bibnamefont {Walter}}, \bibinfo {author} {\bibfnamefont {Z.}~\bibnamefont {Zhu}}, \bibinfo {author} {\bibfnamefont {M.}~\bibnamefont {Gächter}}, \bibinfo {author} {\bibfnamefont {J.}~\bibnamefont {Minguzzi}}, \bibinfo {author} {\bibfnamefont {S.}~\bibnamefont {Roschinski}}, \bibinfo {author} {\bibfnamefont {K.}~\bibnamefont {Sandholzer}}, \bibinfo {author} {\bibfnamefont {K.}~\bibnamefont {Viebahn}}, \ and\ \bibinfo {author} {\bibfnamefont {T.}~\bibnamefont {Esslinger}},\ }\href {\doibase 10.1038/s41567-023-02145-w} {\bibfield  {journal} {\bibinfo  {journal} {Nature Physics}\ }\textbf {\bibinfo {volume} {19}},\ \bibinfo {pages} {1471} (\bibinfo {year} {2023})}\BibitemShut {NoStop}%
\bibitem [{\citenamefont {Nakajima}\ \emph {et~al.}(2021)\citenamefont {Nakajima}, \citenamefont {Takei}, \citenamefont {Sakuma}, \citenamefont {Kuno}, \citenamefont {Marra},\ and\ \citenamefont {Takahashi}}]{nakajima_competition_2021}%
  \BibitemOpen
  \bibfield  {author} {\bibinfo {author} {\bibfnamefont {S.}~\bibnamefont {Nakajima}}, \bibinfo {author} {\bibfnamefont {N.}~\bibnamefont {Takei}}, \bibinfo {author} {\bibfnamefont {K.}~\bibnamefont {Sakuma}}, \bibinfo {author} {\bibfnamefont {Y.}~\bibnamefont {Kuno}}, \bibinfo {author} {\bibfnamefont {P.}~\bibnamefont {Marra}}, \ and\ \bibinfo {author} {\bibfnamefont {Y.}~\bibnamefont {Takahashi}},\ }\href {\doibase 10.1038/s41567-021-01229-9} {\bibfield  {journal} {\bibinfo  {journal} {Nature Physics}\ }\textbf {\bibinfo {volume} {17}},\ \bibinfo {pages} {844} (\bibinfo {year} {2021})}\BibitemShut {NoStop}%
\bibitem [{\citenamefont {Jürgensen}\ \emph {et~al.}(2023)\citenamefont {Jürgensen}, \citenamefont {Mukherjee}, \citenamefont {Jörg},\ and\ \citenamefont {Rechtsman}}]{jurgensen_quantized_2023}%
  \BibitemOpen
  \bibfield  {author} {\bibinfo {author} {\bibfnamefont {M.}~\bibnamefont {Jürgensen}}, \bibinfo {author} {\bibfnamefont {S.}~\bibnamefont {Mukherjee}}, \bibinfo {author} {\bibfnamefont {C.}~\bibnamefont {Jörg}}, \ and\ \bibinfo {author} {\bibfnamefont {M.~C.}\ \bibnamefont {Rechtsman}},\ }\href {\doibase 10.1038/s41567-022-01871-x} {\bibfield  {journal} {\bibinfo  {journal} {Nature Physics}\ }\textbf {\bibinfo {volume} {19}},\ \bibinfo {pages} {420} (\bibinfo {year} {2023})}\BibitemShut {NoStop}%
\bibitem [{\citenamefont {Liu}\ \emph {et~al.}(2025)\citenamefont {Liu}, \citenamefont {Zhang}, \citenamefont {Shi}, \citenamefont {Liu}, \citenamefont {Lu}, \citenamefont {Wang}, \citenamefont {Li}, \citenamefont {Li}, \citenamefont {Deng}, \citenamefont {Zhou}, \citenamefont {Liu}, \citenamefont {Zhang}, \citenamefont {Liang}, \citenamefont {Mei}, \citenamefont {Ma}, \citenamefont {Liu}, \citenamefont {Liu}, \citenamefont {Chen}, \citenamefont {Huang}, \citenamefont {Song}, \citenamefont {Zhao}, \citenamefont {Tian}, \citenamefont {Xiang}, \citenamefont {Zheng}, \citenamefont {Nori}, \citenamefont {Xu},\ and\ \citenamefont {Fan}}]{SC_Thouless_2025}%
  \BibitemOpen
  \bibfield  {author} {\bibinfo {author} {\bibfnamefont {Y.}~\bibnamefont {Liu}}, \bibinfo {author} {\bibfnamefont {Y.-R.}\ \bibnamefont {Zhang}}, \bibinfo {author} {\bibfnamefont {Y.-H.}\ \bibnamefont {Shi}}, \bibinfo {author} {\bibfnamefont {T.}~\bibnamefont {Liu}}, \bibinfo {author} {\bibfnamefont {C.}~\bibnamefont {Lu}}, \bibinfo {author} {\bibfnamefont {Y.-Y.}\ \bibnamefont {Wang}}, \bibinfo {author} {\bibfnamefont {H.}~\bibnamefont {Li}}, \bibinfo {author} {\bibfnamefont {T.-M.}\ \bibnamefont {Li}}, \bibinfo {author} {\bibfnamefont {C.-L.}\ \bibnamefont {Deng}}, \bibinfo {author} {\bibfnamefont {S.-Y.}\ \bibnamefont {Zhou}}, \bibinfo {author} {\bibfnamefont {T.}~\bibnamefont {Liu}}, \bibinfo {author} {\bibfnamefont {J.-C.}\ \bibnamefont {Zhang}}, \bibinfo {author} {\bibfnamefont {G.-H.}\ \bibnamefont {Liang}}, \bibinfo {author} {\bibfnamefont {Z.-Y.}\ \bibnamefont {Mei}}, \bibinfo {author} {\bibfnamefont {W.-G.}\ \bibnamefont {Ma}}, \bibinfo {author} {\bibfnamefont {H.-T.}\ \bibnamefont {Liu}}, \bibinfo
  {author} {\bibfnamefont {Z.-H.}\ \bibnamefont {Liu}}, \bibinfo {author} {\bibfnamefont {C.-T.}\ \bibnamefont {Chen}}, \bibinfo {author} {\bibfnamefont {K.}~\bibnamefont {Huang}}, \bibinfo {author} {\bibfnamefont {X.}~\bibnamefont {Song}}, \bibinfo {author} {\bibfnamefont {S.~P.}\ \bibnamefont {Zhao}}, \bibinfo {author} {\bibfnamefont {Y.}~\bibnamefont {Tian}}, \bibinfo {author} {\bibfnamefont {Z.}~\bibnamefont {Xiang}}, \bibinfo {author} {\bibfnamefont {D.}~\bibnamefont {Zheng}}, \bibinfo {author} {\bibfnamefont {F.}~\bibnamefont {Nori}}, \bibinfo {author} {\bibfnamefont {K.}~\bibnamefont {Xu}}, \ and\ \bibinfo {author} {\bibfnamefont {H.}~\bibnamefont {Fan}},\ }\href {\doibase 10.1038/s41467-024-55343-2} {\bibfield  {journal} {\bibinfo  {journal} {Nature Communications}\ }\textbf {\bibinfo {volume} {16}},\ \bibinfo {pages} {108} (\bibinfo {year} {2025})}\BibitemShut {NoStop}%
\bibitem [{\citenamefont {Boyers}\ \emph {et~al.}(2020)\citenamefont {Boyers}, \citenamefont {Crowley}, \citenamefont {Chandran},\ and\ \citenamefont {Sushkov}}]{Boyers2020}%
  \BibitemOpen
  \bibfield  {author} {\bibinfo {author} {\bibfnamefont {E.}~\bibnamefont {Boyers}}, \bibinfo {author} {\bibfnamefont {P.~J.~D.}\ \bibnamefont {Crowley}}, \bibinfo {author} {\bibfnamefont {A.}~\bibnamefont {Chandran}}, \ and\ \bibinfo {author} {\bibfnamefont {A.~O.}\ \bibnamefont {Sushkov}},\ }\href {\doibase 10.1103/PhysRevLett.125.160505} {\bibfield  {journal} {\bibinfo  {journal} {Phys. Rev. Lett.}\ }\textbf {\bibinfo {volume} {125}},\ \bibinfo {pages} {160505} (\bibinfo {year} {2020})}\BibitemShut {NoStop}%
\bibitem [{\citenamefont {Malz}\ and\ \citenamefont {Smith}(2021)}]{cQED_two_classical_drives_Malz}%
  \BibitemOpen
  \bibfield  {author} {\bibinfo {author} {\bibfnamefont {D.}~\bibnamefont {Malz}}\ and\ \bibinfo {author} {\bibfnamefont {A.}~\bibnamefont {Smith}},\ }\href {\doibase 10.1103/PhysRevLett.126.163602} {\bibfield  {journal} {\bibinfo  {journal} {Phys. Rev. Lett.}\ }\textbf {\bibinfo {volume} {126}},\ \bibinfo {pages} {163602} (\bibinfo {year} {2021})}\BibitemShut {NoStop}%
\bibitem [{\citenamefont {Blais}\ \emph {et~al.}(2021)\citenamefont {Blais}, \citenamefont {Grimsmo}, \citenamefont {Girvin},\ and\ \citenamefont {Wallraff}}]{Blais_review}%
  \BibitemOpen
  \bibfield  {author} {\bibinfo {author} {\bibfnamefont {A.}~\bibnamefont {Blais}}, \bibinfo {author} {\bibfnamefont {A.~L.}\ \bibnamefont {Grimsmo}}, \bibinfo {author} {\bibfnamefont {S.~M.}\ \bibnamefont {Girvin}}, \ and\ \bibinfo {author} {\bibfnamefont {A.}~\bibnamefont {Wallraff}},\ }\href {http://dx.doi.org/10.1103/RevModPhys.93.025005} {\bibfield  {journal} {\bibinfo  {journal} {Reviews of Modern Physics}\ }\textbf {\bibinfo {volume} {93}} (\bibinfo {year} {2021})}\BibitemShut {NoStop}%
\bibitem [{\citenamefont {Krantz}\ \emph {et~al.}(2019)\citenamefont {Krantz}, \citenamefont {Kjaergaard}, \citenamefont {Yan}, \citenamefont {Orlando}, \citenamefont {Gustavsson},\ and\ \citenamefont {Oliver}}]{quantum_engineer_guide}%
  \BibitemOpen
  \bibfield  {author} {\bibinfo {author} {\bibfnamefont {P.}~\bibnamefont {Krantz}}, \bibinfo {author} {\bibfnamefont {M.}~\bibnamefont {Kjaergaard}}, \bibinfo {author} {\bibfnamefont {F.}~\bibnamefont {Yan}}, \bibinfo {author} {\bibfnamefont {T.~P.}\ \bibnamefont {Orlando}}, \bibinfo {author} {\bibfnamefont {S.}~\bibnamefont {Gustavsson}}, \ and\ \bibinfo {author} {\bibfnamefont {W.~D.}\ \bibnamefont {Oliver}},\ }\href {\doibase 10.1063/1.5089550} {\bibfield  {journal} {\bibinfo  {journal} {Applied Physics Reviews}\ }\textbf {\bibinfo {volume} {6}},\ \bibinfo {pages} {021318} (\bibinfo {year} {2019})}\BibitemShut {NoStop}%
\bibitem [{\citenamefont {Ritter}\ \emph {et~al.}(2024)\citenamefont {Ritter}, \citenamefont {Long}, \citenamefont {Yue}, \citenamefont {Chandran},\ and\ \citenamefont {Kollár}}]{dissipation_paper}%
  \BibitemOpen
  \bibfield  {author} {\bibinfo {author} {\bibfnamefont {M.}~\bibnamefont {Ritter}}, \bibinfo {author} {\bibfnamefont {D.~M.}\ \bibnamefont {Long}}, \bibinfo {author} {\bibfnamefont {Q.}~\bibnamefont {Yue}}, \bibinfo {author} {\bibfnamefont {A.}~\bibnamefont {Chandran}}, \ and\ \bibinfo {author} {\bibfnamefont {A.~J.}\ \bibnamefont {Kollár}},\ }\href {https://arxiv.org/abs/2410.12908} {\enquote {\bibinfo {title} {Autonomous stabilization of floquet states using static dissipation},}\ } (\bibinfo {year} {2024}),\ \Eprint {http://arxiv.org/abs/2410.12908}{arXiv:2410.12908 [quant-ph]}\BibitemShut {NoStop}%
\bibitem [{\citenamefont {Kolodrubetz}\ \emph {et~al.}(2018)\citenamefont {Kolodrubetz}, \citenamefont {Nathan}, \citenamefont {Gazit}, \citenamefont {Morimoto},\ and\ \citenamefont {Moore}}]{Kolodrubetz2018}%
  \BibitemOpen
  \bibfield  {author} {\bibinfo {author} {\bibfnamefont {M.~H.}\ \bibnamefont {Kolodrubetz}}, \bibinfo {author} {\bibfnamefont {F.}~\bibnamefont {Nathan}}, \bibinfo {author} {\bibfnamefont {S.}~\bibnamefont {Gazit}}, \bibinfo {author} {\bibfnamefont {T.}~\bibnamefont {Morimoto}}, \ and\ \bibinfo {author} {\bibfnamefont {J.~E.}\ \bibnamefont {Moore}},\ }\href@noop {} {\bibfield  {journal} {\bibinfo  {journal} {Phys. Rev. Lett.}\ }\textbf {\bibinfo {volume} {120}},\ \bibinfo {pages} {150601} (\bibinfo {year} {2018})}\BibitemShut {NoStop}%
\bibitem [{\citenamefont {Crowley}\ \emph {et~al.}(2020)\citenamefont {Crowley}, \citenamefont {Martin},\ and\ \citenamefont {Chandran}}]{Crowley:2020tl}%
  \BibitemOpen
  \bibfield  {author} {\bibinfo {author} {\bibfnamefont {P.~J.~D.}\ \bibnamefont {Crowley}}, \bibinfo {author} {\bibfnamefont {I.}~\bibnamefont {Martin}}, \ and\ \bibinfo {author} {\bibfnamefont {A.}~\bibnamefont {Chandran}},\ }\href@noop {} {\bibfield  {journal} {\bibinfo  {journal} {Phys. Rev. Lett.}\ }\textbf {\bibinfo {volume} {125}},\ \bibinfo {pages} {100601} (\bibinfo {year} {2020})}\BibitemShut {NoStop}%
\bibitem [{\citenamefont {Giovannetti}\ \emph {et~al.}(2011)\citenamefont {Giovannetti}, \citenamefont {Lloyd},\ and\ \citenamefont {Maccone}}]{Giovannetti2011metrology}%
  \BibitemOpen
  \bibfield  {author} {\bibinfo {author} {\bibfnamefont {V.}~\bibnamefont {Giovannetti}}, \bibinfo {author} {\bibfnamefont {S.}~\bibnamefont {Lloyd}}, \ and\ \bibinfo {author} {\bibfnamefont {L.}~\bibnamefont {Maccone}},\ }\href {\doibase 10.1038/nphoton.2011.35} {\bibfield  {journal} {\bibinfo  {journal} {Nature Photonics}\ }\textbf {\bibinfo {volume} {5}},\ \bibinfo {pages} {222} (\bibinfo {year} {2011})}\BibitemShut {NoStop}%
\bibitem [{\citenamefont {Citro}\ and\ \citenamefont {Aidelsburger}(2023{\natexlab{b}})}]{Citro2023pumpreview}%
  \BibitemOpen
  \bibfield  {author} {\bibinfo {author} {\bibfnamefont {R.}~\bibnamefont {Citro}}\ and\ \bibinfo {author} {\bibfnamefont {M.}~\bibnamefont {Aidelsburger}},\ }\href {\doibase 10.1038/s42254-022-00545-0} {\bibfield  {journal} {\bibinfo  {journal} {Nature Reviews Physics}\ }\textbf {\bibinfo {volume} {5}},\ \bibinfo {pages} {87} (\bibinfo {year} {2023}{\natexlab{b}})}\BibitemShut {NoStop}%
\bibitem [{\citenamefont {Ozawa}\ and\ \citenamefont {Price}(2019)}]{Ozawa2019syntheticreview}%
  \BibitemOpen
  \bibfield  {author} {\bibinfo {author} {\bibfnamefont {T.}~\bibnamefont {Ozawa}}\ and\ \bibinfo {author} {\bibfnamefont {H.~M.}\ \bibnamefont {Price}},\ }\href {\doibase 10.1038/s42254-019-0045-3} {\bibfield  {journal} {\bibinfo  {journal} {Nature Reviews Physics}\ }\textbf {\bibinfo {volume} {1}},\ \bibinfo {pages} {349} (\bibinfo {year} {2019})}\BibitemShut {NoStop}%
\bibitem [{\citenamefont {Sambe}(1973)}]{Sambe1973synthetic}%
  \BibitemOpen
  \bibfield  {author} {\bibinfo {author} {\bibfnamefont {H.}~\bibnamefont {Sambe}},\ }\href {\doibase 10.1103/PhysRevA.7.2203} {\bibfield  {journal} {\bibinfo  {journal} {Phys. Rev. A}\ }\textbf {\bibinfo {volume} {7}},\ \bibinfo {pages} {2203} (\bibinfo {year} {1973})}\BibitemShut {NoStop}%
\bibitem [{\citenamefont {Ho}\ \emph {et~al.}(1983)\citenamefont {Ho}, \citenamefont {Chu},\ and\ \citenamefont {Tietz}}]{Ho1983}%
  \BibitemOpen
  \bibfield  {author} {\bibinfo {author} {\bibfnamefont {T.-S.}\ \bibnamefont {Ho}}, \bibinfo {author} {\bibfnamefont {S.-I.}\ \bibnamefont {Chu}}, \ and\ \bibinfo {author} {\bibfnamefont {J.~V.}\ \bibnamefont {Tietz}},\ }\href@noop {} {\bibfield  {journal} {\bibinfo  {journal} {Chem. Phys. Lett.}\ }\textbf {\bibinfo {volume} {96}},\ \bibinfo {pages} {464} (\bibinfo {year} {1983})}\BibitemShut {NoStop}%
\bibitem [{\citenamefont {Verdeny}\ \emph {et~al.}(2016)\citenamefont {Verdeny}, \citenamefont {Puig},\ and\ \citenamefont {Mintert}}]{Verdeny_2016_synthetic_dim}%
  \BibitemOpen
  \bibfield  {author} {\bibinfo {author} {\bibfnamefont {A.}~\bibnamefont {Verdeny}}, \bibinfo {author} {\bibfnamefont {J.}~\bibnamefont {Puig}}, \ and\ \bibinfo {author} {\bibfnamefont {F.}~\bibnamefont {Mintert}},\ }\href {\doibase 10.1515/zna-2016-0079} {\bibfield  {journal} {\bibinfo  {journal} {Zeitschrift für Naturforschung A}\ }\textbf {\bibinfo {volume} {71}},\ \bibinfo {pages} {897} (\bibinfo {year} {2016})}\BibitemShut {NoStop}%
\bibitem [{\citenamefont {Koch}\ \emph {et~al.}(2007)\citenamefont {Koch}, \citenamefont {Yu}, \citenamefont {Gambetta}, \citenamefont {Houck}, \citenamefont {Schuster}, \citenamefont {Majer}, \citenamefont {Blais}, \citenamefont {Devoret}, \citenamefont {Girvin},\ and\ \citenamefont {Schoelkopf}}]{Koch2007transmon}%
  \BibitemOpen
  \bibfield  {author} {\bibinfo {author} {\bibfnamefont {J.}~\bibnamefont {Koch}}, \bibinfo {author} {\bibfnamefont {T.~M.}\ \bibnamefont {Yu}}, \bibinfo {author} {\bibfnamefont {J.}~\bibnamefont {Gambetta}}, \bibinfo {author} {\bibfnamefont {A.~A.}\ \bibnamefont {Houck}}, \bibinfo {author} {\bibfnamefont {D.~I.}\ \bibnamefont {Schuster}}, \bibinfo {author} {\bibfnamefont {J.}~\bibnamefont {Majer}}, \bibinfo {author} {\bibfnamefont {A.}~\bibnamefont {Blais}}, \bibinfo {author} {\bibfnamefont {M.~H.}\ \bibnamefont {Devoret}}, \bibinfo {author} {\bibfnamefont {S.~M.}\ \bibnamefont {Girvin}}, \ and\ \bibinfo {author} {\bibfnamefont {R.~J.}\ \bibnamefont {Schoelkopf}},\ }\href {\doibase 10.1103/PhysRevA.76.042319} {\bibfield  {journal} {\bibinfo  {journal} {Phys. Rev. A}\ }\textbf {\bibinfo {volume} {76}},\ \bibinfo {pages} {042319} (\bibinfo {year} {2007})}\BibitemShut {NoStop}%
\bibitem [{\citenamefont {Schreier}\ \emph {et~al.}(2008)\citenamefont {Schreier}, \citenamefont {Houck}, \citenamefont {Koch}, \citenamefont {Schuster}, \citenamefont {Johnson}, \citenamefont {Chow}, \citenamefont {Gambetta}, \citenamefont {Majer}, \citenamefont {Frunzio}, \citenamefont {Devoret}, \citenamefont {Girvin},\ and\ \citenamefont {Schoelkopf}}]{Schreier2008transmon}%
  \BibitemOpen
  \bibfield  {author} {\bibinfo {author} {\bibfnamefont {J.~A.}\ \bibnamefont {Schreier}}, \bibinfo {author} {\bibfnamefont {A.~A.}\ \bibnamefont {Houck}}, \bibinfo {author} {\bibfnamefont {J.}~\bibnamefont {Koch}}, \bibinfo {author} {\bibfnamefont {D.~I.}\ \bibnamefont {Schuster}}, \bibinfo {author} {\bibfnamefont {B.~R.}\ \bibnamefont {Johnson}}, \bibinfo {author} {\bibfnamefont {J.~M.}\ \bibnamefont {Chow}}, \bibinfo {author} {\bibfnamefont {J.~M.}\ \bibnamefont {Gambetta}}, \bibinfo {author} {\bibfnamefont {J.}~\bibnamefont {Majer}}, \bibinfo {author} {\bibfnamefont {L.}~\bibnamefont {Frunzio}}, \bibinfo {author} {\bibfnamefont {M.~H.}\ \bibnamefont {Devoret}}, \bibinfo {author} {\bibfnamefont {S.~M.}\ \bibnamefont {Girvin}}, \ and\ \bibinfo {author} {\bibfnamefont {R.~J.}\ \bibnamefont {Schoelkopf}},\ }\href {\doibase 10.1103/PhysRevB.77.180502} {\bibfield  {journal} {\bibinfo  {journal} {Phys. Rev. B}\ }\textbf {\bibinfo {volume} {77}},\ \bibinfo {pages} {180502} (\bibinfo {year} {2008})}\BibitemShut
  {NoStop}%
\bibitem [{\citenamefont {Walls}\ and\ \citenamefont {Milburn}(2008)}]{wallsMilburn}%
  \BibitemOpen
  \bibfield  {author} {\bibinfo {author} {\bibfnamefont {D.~F.}\ \bibnamefont {Walls}}\ and\ \bibinfo {author} {\bibfnamefont {G.~J.}\ \bibnamefont {Milburn}},\ }\href@noop {} {\emph {\bibinfo {title} {{Quantum Optics}}}},\ Springer\ (\bibinfo  {publisher} {Springer},\ \bibinfo {year} {2008})\BibitemShut {NoStop}%
\bibitem [{\citenamefont {Foot}(2005)}]{atomic_physics_Foot}%
  \BibitemOpen
  \bibfield  {author} {\bibinfo {author} {\bibfnamefont {C.~J.}\ \bibnamefont {Foot}},\ }\href@noop {} {\emph {\bibinfo {title} {Atomic Physics}}}\ (\bibinfo  {publisher} {Oxford University Press},\ \bibinfo {year} {2005})\BibitemShut {NoStop}%
\bibitem [{\citenamefont {Rol}\ \emph {et~al.}(2020)\citenamefont {Rol}, \citenamefont {Ciorciaro}, \citenamefont {Malinowski}, \citenamefont {Tarasinski}, \citenamefont {Sagastizabal}, \citenamefont {Bultink}, \citenamefont {Salathe}, \citenamefont {Haandbaek}, \citenamefont {Sedivy},\ and\ \citenamefont {DiCarlo}}]{cryoscope}%
  \BibitemOpen
  \bibfield  {author} {\bibinfo {author} {\bibfnamefont {M.~A.}\ \bibnamefont {Rol}}, \bibinfo {author} {\bibfnamefont {L.}~\bibnamefont {Ciorciaro}}, \bibinfo {author} {\bibfnamefont {F.~K.}\ \bibnamefont {Malinowski}}, \bibinfo {author} {\bibfnamefont {B.~M.}\ \bibnamefont {Tarasinski}}, \bibinfo {author} {\bibfnamefont {R.~E.}\ \bibnamefont {Sagastizabal}}, \bibinfo {author} {\bibfnamefont {C.~C.}\ \bibnamefont {Bultink}}, \bibinfo {author} {\bibfnamefont {Y.}~\bibnamefont {Salathe}}, \bibinfo {author} {\bibfnamefont {N.}~\bibnamefont {Haandbaek}}, \bibinfo {author} {\bibfnamefont {J.}~\bibnamefont {Sedivy}}, \ and\ \bibinfo {author} {\bibfnamefont {L.}~\bibnamefont {DiCarlo}},\ }\href {\doibase 10.1063/1.5133894} {\bibfield  {journal} {\bibinfo  {journal} {Applied Physics Letters}\ }\textbf {\bibinfo {volume} {116}},\ \bibinfo {pages} {054001} (\bibinfo {year} {2020})}\BibitemShut {NoStop}%
\bibitem [{\citenamefont {Cohen-Tannoudji}\ \emph {et~al.}(1998)\citenamefont {Cohen-Tannoudji}, \citenamefont {Dupont-Roc},\ and\ \citenamefont {Grynberg}}]{CohenTannoudji}%
  \BibitemOpen
  \bibfield  {author} {\bibinfo {author} {\bibfnamefont {C.}~\bibnamefont {Cohen-Tannoudji}}, \bibinfo {author} {\bibfnamefont {J.}~\bibnamefont {Dupont-Roc}}, \ and\ \bibinfo {author} {\bibfnamefont {G.}~\bibnamefont {Grynberg}},\ }\href@noop {} {\emph {\bibinfo {title} {Atom—Photon Interactions: Basic Process and Appilcations}}}\ (\bibinfo  {publisher} {Wiley},\ \bibinfo {year} {1998})\BibitemShut {NoStop}%
\bibitem [{\citenamefont {Ithier}\ \emph {et~al.}(2005)\citenamefont {Ithier}, \citenamefont {Collin}, \citenamefont {Joyez}, \citenamefont {Meeson}, \citenamefont {Vion}, \citenamefont {Esteve}, \citenamefont {Chiarello}, \citenamefont {Shnirman}, \citenamefont {Makhlin}, \citenamefont {Schriefl} \emph {et~al.}}]{ithier2005decoherence}%
  \BibitemOpen
  \bibfield  {author} {\bibinfo {author} {\bibfnamefont {G.}~\bibnamefont {Ithier}}, \bibinfo {author} {\bibfnamefont {E.}~\bibnamefont {Collin}}, \bibinfo {author} {\bibfnamefont {P.}~\bibnamefont {Joyez}}, \bibinfo {author} {\bibfnamefont {P.}~\bibnamefont {Meeson}}, \bibinfo {author} {\bibfnamefont {D.}~\bibnamefont {Vion}}, \bibinfo {author} {\bibfnamefont {D.}~\bibnamefont {Esteve}}, \bibinfo {author} {\bibfnamefont {F.}~\bibnamefont {Chiarello}}, \bibinfo {author} {\bibfnamefont {A.}~\bibnamefont {Shnirman}}, \bibinfo {author} {\bibfnamefont {Y.}~\bibnamefont {Makhlin}}, \bibinfo {author} {\bibfnamefont {J.}~\bibnamefont {Schriefl}},  \emph {et~al.},\ }\href@noop {} {\bibfield  {journal} {\bibinfo  {journal} {Physical Review B—Condensed Matter and Materials Physics}\ }\textbf {\bibinfo {volume} {72}},\ \bibinfo {pages} {134519} (\bibinfo {year} {2005})}\BibitemShut {NoStop}%
\bibitem [{\citenamefont {Slichter}(1990)}]{slichter_NMR}%
  \BibitemOpen
  \bibfield  {author} {\bibinfo {author} {\bibfnamefont {C.}~\bibnamefont {Slichter}},\ }\href@noop {} {\emph {\bibinfo {title} {Principles of Magnetic Resonance}}}\ (\bibinfo  {publisher} {Springer},\ \bibinfo {year} {1990})\BibitemShut {NoStop}%
\end{thebibliography}%

\clearpage
\newpage
\onecolumngrid
\appendix

\clearpage
\newpage
\section{Time dynamics and Fourier spectrum of adiabatic following data}\label{appendix:supp_data}

\begin{figure*}[ht]
    \centering
    \includegraphics[width=\linewidth]{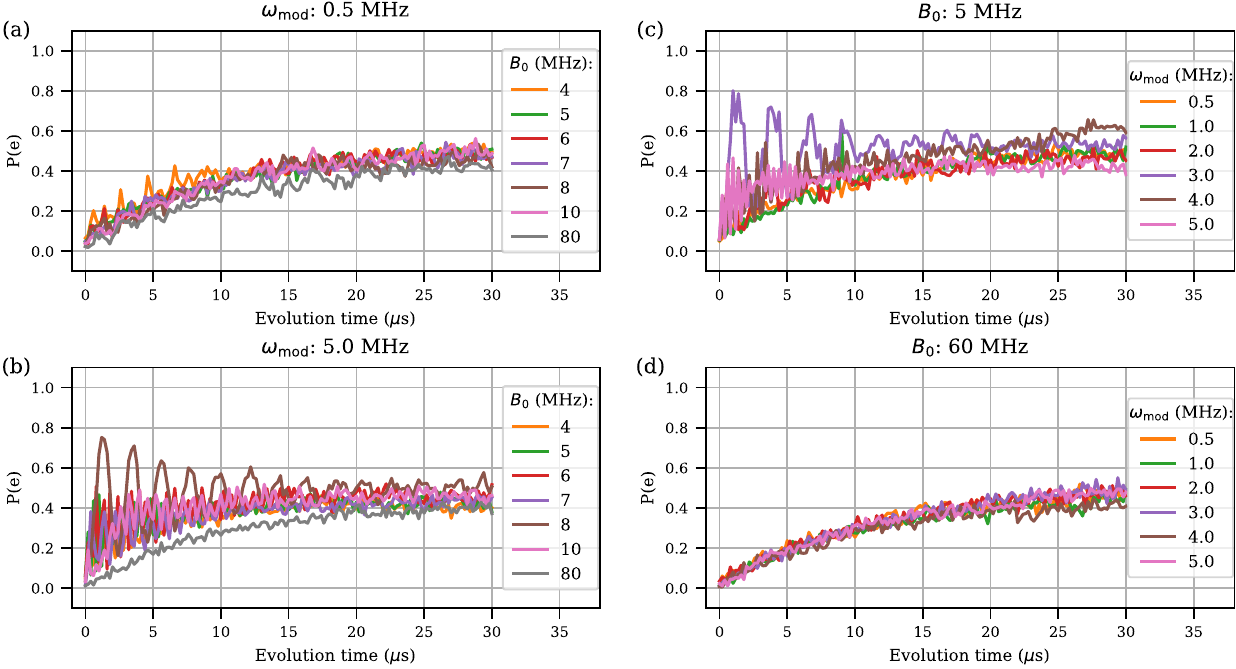}
    \caption{Adiabatic basis measurements of the qubit under the rotating field versus the field strength $B_0$ and the rotation rate $\omega_{\mathrm{mod}}$. In all cases, the qubit is prepared in the laboratory-frame ground state and this initial state is adiabatically mapped back to itself after applying the  rotating field Hamiltonian.
    (a)-(b) Each panel shows the probability of excitation versus time for variable $B_0$ and fixed $\omega_{\mathrm{mod}}$. (a) For slow rotation rates ($\omega_{\mathrm{mod}}=0.5$ MHz), the qubit follows the applied field for all measured field strengths $B_0$ and decays exponentially to a mixed state with a decay rate independent of $B_0$. (b) For fast rotation rates ($\omega_{\mathrm{mod}}=5$ MHz), the breakdown in adiabatic following is visible in the emergence of large contrast oscillations superposed on the exponential decay. These features correspond to nutation about $\vec{B}(t)$. For $\omega_{\mathrm{mod}}=5$ MHz, adiabatic following occurs when the field strength is larger than the threshold $B_0=10$ MHz. (c)-(d) Each panel shows the probability of excitation versus time for variable $\omega_{\mathrm{mod}}$ and fixed $B_0$. (c) For small field strengths ($B_0=5$ MHz), adiabatic following breaks down as the rotation rate increases. For $B_0=5$ MHz, adiabatic following occurs when the rotation frequency is smaller than the threshold $\omega_{\mathrm{mod}} = 1$ MHz. (d) For large field strengths ($B_0=60$ MHz), the qubit adiabatically follows the applied field for all measured rotation rates $\omega_{\mathrm{mod}}$ and decays exponentially to a mixed state with a decay rate independent of $\omega_{\mathrm{mod}}$. 
    }
    
    \label{fig:slow_sweep_data}
\end{figure*}

\begin{figure*}[ht]
    \centering
    \includegraphics[width=\linewidth]{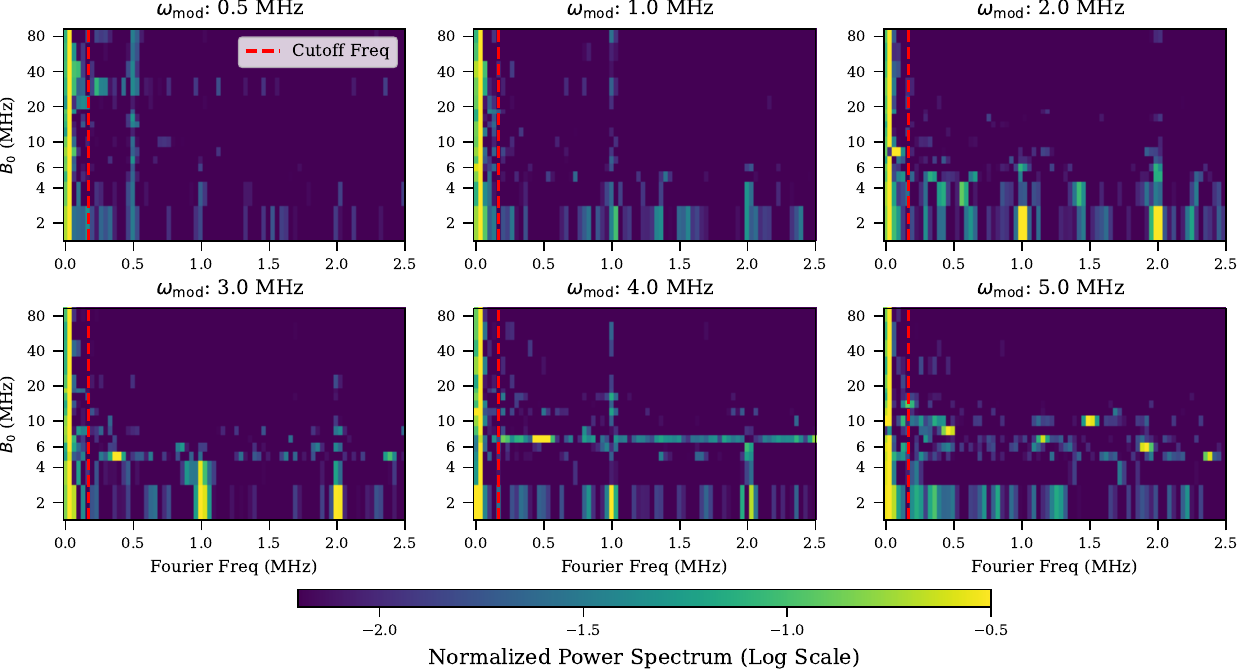}
    \caption{Normalized power spectrum of the adiabatic basis qubit dynamics versus rotation rate $\omega_{\mathrm{mod}}$  and field strength $B_0$. 
    (Example time traces are shown in Fig.~\ref{fig:slow_sweep_data}.)
    Each panel shows the normalized power spectrum for variable $B_0$ and fixed $\omega_{\mathrm{mod}}$. In the case of ideal adiabatic following, the Fourier spectrum is a Lorentzian centered at zero frequency. Adiabatic breakdown is then identified as the emergence of peaks at higher frequencies or by a general increase in the high-frequency portion of the spectrum. Good adiabatic following is clearly visible for $B_0\gtrsim 2.5~\omega_{\mathrm{mod}}$ as the peak Fourier content is centered at zero frequency with minimal high frequency noise. The red dashed line shows the location of the cutoff used to compute the fractional harmonic content defined in Eq.~\eqref{eqn:metric} of the main text. Fig.~\ref{fig:adiabatic_sweep} in the main text shows the fractional harmonic content of this data versus $\omega_{\mathrm{mod}}$ and $B_0$.}
    \label{fig:slow_sweep_FFT}
\end{figure*}

\begin{figure*}[ht]
    \centering
    \includegraphics[width=\linewidth]{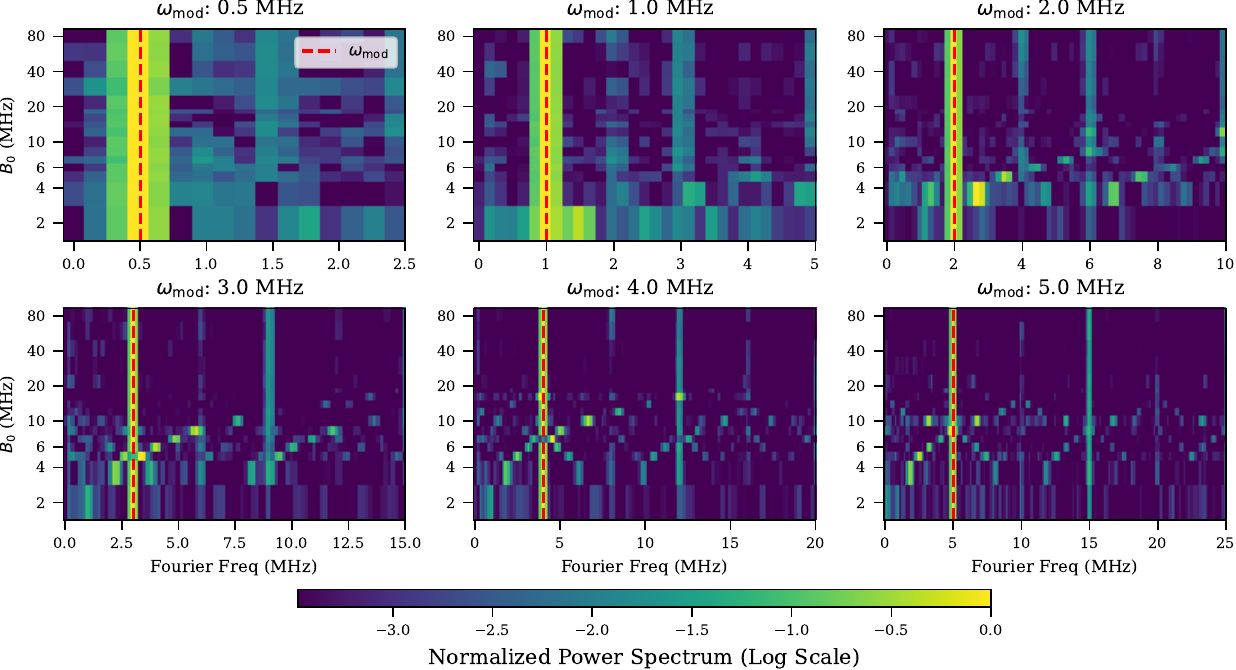}
    \caption{Normalized power spectrum of the bare-basis qubit dynamics versus rotation rate $\omega_{\mathrm{mod}}$  and field strength $B_0$. 
    (Example time traces are shown in Fig.~\ref{fig:adiabatic_sweep} (a)-(b).) 
    Each panel shows the normalized power spectrum for variable $B_0$ and fixed $\omega_{\mathrm{mod}}$ [red dashed line indicates $\omega_{\mathrm{mod}}$]. With ideal adiabatic following, the bare-basis measurement should be an exponentially decaying sinusoidal function with frequency $\omega_{\mathrm{mod}}$. This corresponds to a large Fourier amplitude at $\omega_{\mathrm{mod}}$ (and its harmonics) but low amplitudes at other frequencies. Breakdown of adiabatic following is visible as additional features appearing at frequencies other than $\omega_{\mathrm{mod}}$(or its harmonics). The additional features are due to nutation about the applied field and scale linearly with $B_0$. Note that the linearity is not clearly visible in each panel due to logarithmic axes. Similar to Fig.~\ref{fig:slow_sweep_FFT}, good adiabatic following is clearly visible for $B_0\gtrsim 2.5~\omega_{\mathrm{mod}}$ when the additional Fourier spurs disappear.}
    \label{fig:fast_sweep_FFT}
\end{figure*}

In this section we discuss the breakdown of the adiabatic following at small values of the ratio $B_0/\omega_{\mathrm{mod}}$, as measured in both the adiabatic basis and the bare basis. To quantify this, we examine the Fourier spectrum of the data in Fig.~\ref{fig:slow_sweep_data} and show that the breakdown is accompanied by the emergence of additional frequency components. 

We first show the breakdown of adiabatic following for measurements in the adiabatic basis of the rotating field. In this measurement, the instantaneous eigenstates with spin projections $\pm1$ along the rotating field are mapped onto $\ket{g}$,$\ket{e}$ before readout. For ideal following, the qubit follows the applied field perfectly until incoherent external processes cause the adiabatic basis population to decay on the timescale $T_d$ (see Fig.~\ref{fig:time_scale_comp} in the main text). For sufficiently large rotation rates (or small applied field strengths), the qubit does not follow the applied field and instead it nutates about the applied field. The transition from adiabatic following to nutation is visible in Fig.~\ref{fig:slow_sweep_data}. For very slow rotation rates or large field strengths, the qubit follows the applied field independent of the field amplitude or rotation rate respectively. At higher rotation rates or smaller applied fields, the breakdown in adiabatic following is visible in the emergence of an oscillatory component in the traces. 

The breakdown of adiabatic following is more striking in the Fourier transform of the data. For ideal following, the exponential decay leads to a Lorentzian response centered at zero frequency, with a width set by the decay rate. The breakdown regime is identified as deviations from this Lorentzian behavior, either through the emergence of peaks at higher frequencies or by a general increase in the high-frequency portion of the spectrum. The power spectrum for different rotation rates and field strengths is shown in Fig.~\ref{fig:slow_sweep_FFT}. The adiabatic following regime is clearly identifiable for $B_0\gtrsim ~2.5~\omega_{\mathrm{mod}}$, as the region in which the data exhibits low Fourier amplitude at higher frequencies. As stated in the main text [Eq.~\eqref{eqn:metric}], we quantify the breakdown of adiabatic following using the fraction of the spectrum above the cutoff frequency (defined as the width of the ideal Lorentzian response and shown as a dashed red line in Fig.~\ref{fig:slow_sweep_FFT}). The value of each pixel in Fig.~\ref{fig:adiabatic_sweep}(c) of the main text is the fraction of power in the spectrum above this cutoff frequency. 

The breakdown of adiabatic following is also visible in the bare qubit basis measurements. In bare basis measurements, ideal following is characterized by oscillations at the field rotation frequency $\omega_{\mathrm{mod}}$. As with the adiabatic basis measurements, the oscillations damp out to a mixed state as the qubit depolarizes. 
When the Fourier spectrum only exhibits peaks at $\omega_{\mathrm{mod}}$ and its harmonics, the qubit is successfully following the applied field. The power spectrum versus Fourier frequency at different $\omega_{\mathrm{mod}}$ and $B_0$ are shown in Fig.~\ref{fig:fast_sweep_FFT}.

\section{Flux bias line filter compensation}\label{appendix:precompensation}

\begin{figure*}[ht]
\centering
    \includegraphics[width=\textwidth]{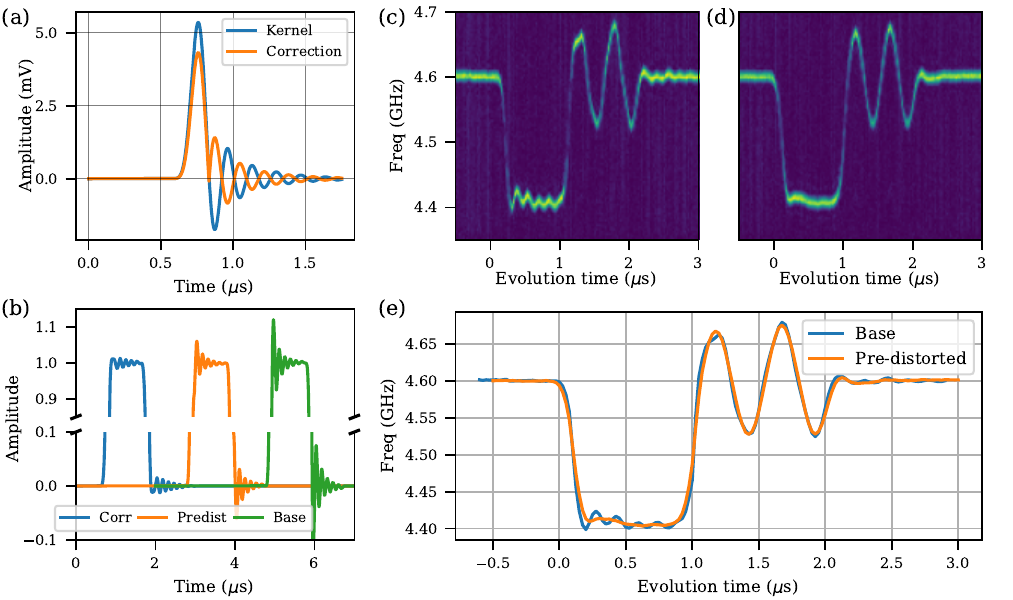}
    \vspace{-0.2cm}
    \caption{\label{fig:precompensation} 
    Filter compensation: (a) Measured filter convolution kernel (blue) acquired from the delta function response of an SLP5+ filter at room temperature. The compensation kernel (orange) is computed by flipping the sign of the oscillatory portion. Both kernels are normalized to provide unity gain. (b) Numerical simulation of corrected (blue) and uncorrected (green) square pulses going through the filter showing a 5x reduction in ringing amplitude. The compensation kernel is used to generate the predistorted trace (orange) which is shown \emph{before} going through the filter. (c)-(d) Qubit spectroscopy during the full $z$ protocol without (c) and with (d) pulse predistortion. Ripples after the initial ramp at \(t=0\) are reduced and the amplitude of the oscillatory portion of the schedule becomes more uniform. (e) Qubit frequency extracted from panels (c) and (d). 
    } 
\end{figure*}

In this section, we present a precompensation technique which mitigates ringing that occurs on the flux bias line due to the presence of a low-pass filter on the line needed to protect the qubit from environmental noise. This ringing introduces errors in the flux pulse shape applied to the qubit. When uncorrected, it prevents accurate measurements of $\sigma_x$ at the end of an experiment since the qubit frequency oscillates during the $\pi$/2 pulse due to the sudden shutoff of the $z$-component of the field.

We use a low-pass filter (minicircuits SLP 5+) on the on-chip flux bias line to reduce the bandwidth of the noise reaching the qubit, increasing its coherence time. To maximize the available bandwidth but minimize the noise power reaching the qubit, the filter is a sharp elliptical filter with a cutoff frequency at 6 MHz. The filter introduces two artifacts that must be corrected: a group delay which increases drastically near the cutoff (from 140 ns below 4 MHz to 320 ns near 6 MHz) and a ringing response due to the sharp cutoff. The group delay is calibrated out with the method defined in the main text and assumed to be a constant over the range of frequencies considered ($\omega_{\mathrm{mod}}=$ 1 to 5 MHz). To reduce the ringing in the FBL pulses reaching the qubit, we perform a simplified version of compensation which pre-distorts the signal before reaching the filter in such a way as to minimize the ringing response of the filter. 

The required compensation is calculated from the time-domain response of the filter. The response is acquired at room temperature by sending a 10 ns pulse to the filter and measuring the ringing response.
This response in shown in Fig.~\ref{fig:precompensation}(a) (blue) and displays two types of features: a primary peak acting as smoothing function on the data and damped oscillatory response which introduces the ringing in the filtered data. 
Our approach to reduce the ringing is to simply flip the sign of the ringing response. The resulting precompensation filter effectively introduces ripples of the opposite phase in the data and is shown in orange in Fig.~\ref{fig:precompensation}(a).
We simulate the effect of the filter on a square pulse numerically, shown in Fig.~\ref{fig:precompensation}(b). The uncorrected response (green) displays large ringing at the edges of the pulse whereas the computed predistortion (orange) displays oscillations of the opposite phase. The corrected response (blue), which is the combination of the two, shows a factor of 5 reduction in the ripple amplitude. 
While more sophisticated methods exist to predistort the signal \cite{cryoscope}, this method is computationally simple to apply and provides a large enough ripple suppression for the proof of concept measurements presented here. Finally, the predistortion kernel introduces some damping near the cutoff frequency. This is compensated for by increasing the amplitude of the oscillatory frequency portion of the applied $z$-field. 

The effect of the precompensation is also verified with qubit spectroscopy with and without predistortion, shown in Fig.~\ref{fig:precompensation}(c)-(d).
The time dynamics of the FBL are acquired by performing pulsed spectroscopy at each time step with a short pulse (50 ns) to determine the real-time qubit frequency. While the finite duration of the spectroscopy pulse leads to some smearing of the time domain data, the pulses cannot be shortened further without the frequency resolution becoming too low.
The effect of the predistortion is visible in the extracted qubit frequency [Fig.~\ref{fig:precompensation}(e)] as a reduction in ringing near the sharp transitions in the FBL pulse as well as a smoother transition to the oscillatory portion of the driving protocol. 
This spectroscopy technique is a simplified version of the cryoscope method described in Ref. \cite{cryoscope} in which a Ramsey-type measurement is used to measure the accrued phase for fast flux pulses.  

\section{State initialization and measurement}\label{appendix:initialization}

\begin{figure}
    \centering
    \includegraphics[width=0.85\linewidth]{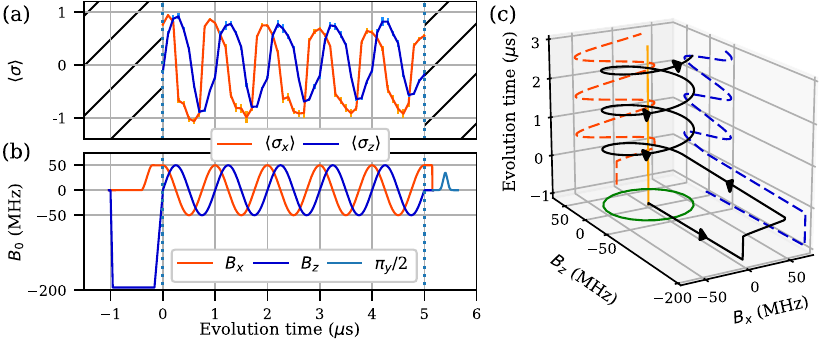}
    \caption{
    $\vec{B}(t)$ and the measured qubit response. 
    (a) Qubit state projections $\langle\sigma_z\rangle$ (blue) and $\langle\sigma_x\rangle$ (orange) onto the Bloch sphere during the rotating field evolution. Qubit expectation values are measured by evolving for a variable total rotation time $<5~\mu$s. These measurements show that the qubit state is aligned with the applied rotating field. 
    (b) Time dependence of $B_x$ (orange) and $B_z$ (blue) for $5~\mu$s of evolution, including the adiabatic preparation portion of the protocol at $t<0$, and the measurement preparation steps at $t>5~\mu$s. The optional $\pi/2$ pulse used to obtain $\langle\sigma_x\rangle$ is shown in light blue.
    (c) 3D visualization of $\vec{B}(t)$, shown in black. The $x$ (orange dashed) and $z$ components (blue dashed) of the field are shown projected onto the back walls of the display. The green circle indicates the circles traced out by $\vec{B}(t)$ during the rotating field evolution, and the yellow line marks the center of this circle at $B_x=B_z=0$. 
    }
    \label{fig:protocol}
\end{figure}
Due to the large field amplitudes, pulsed initialization of the qubit into the state aligned or anti-aligned with the $t=0$ driving field is challenging. Instead, we perform the following ramp-in procedure to adiabatically prepare the desired states \cite{dissipation_paper}: (i) the qubit frequency is shifted by $-\min\{200, 4B_0\}$ MHz from the operating point, (ii) the microwave drive is slowly turned on, and (iii) the qubit is ramped back into resonance with the drive. This initializes the state to lie in the equator of the Bloch sphere. This ramp-in procedure maintains the qubit state pointing along the applied magnetic field and reduces any sudden changes in field angle as the microwave drive is turned on while maintaining a large frequency offset. By prepending an optional $\pi$-pulse before the initialization procedure, the qubit can be prepared in either the $\ket{+}$ or $\ket{-}$ state. The full protocol is shown in Fig.~\ref{fig:protocol}.

Due to the finite bandwidth of the FBL filter (with cutoff frequency of 5 MHz), the $z$-field cannot be shut off instantaneously for the bare basis measurement. Since it is not possible to diabatically turn off both axes of the applied field, we instead implement a staged turn off in which one component is held fixed while the other is shut off as quickly as possible.  While this shutoff procedure lead to precession of the qubit state before readout, crucially, the precession occurs around the \emph{measurement axis}, and does not affect the final state readout. For $z$ measurements, this requires the $z$ field to be kept on for an additional 50 ns while the $x$ field is turned off before ramping to 0 and performing state measurement. For $x$ measurements, the reverse sequence is used.

Adiabatic basis measurements require rotating the field axis back to $-z$ before each readout using the ramp out protocol designed in Ref.~\cite{dissipation_paper}. To ensure adiabatic following at small field sizes and improve performance of state readout, we use a large $x$ field during the adiabatic ramp out. Specifically, (i) we stop rotating the field, (ii) hold $B_z$ constant for a variable buffer time to allow ringing induced by the cusp in $B_z(t)$ to settle, (iii) simultaneously ramp $B_x$ to $\max\{25,B_{0}\}$ MHz, (iv) ramp $B_z$ to $-\min\{200, 4B_0\}$ MHz, (v) ramp down $B_x$, (vi) ramp down $B_z$. A buffer time of 0.15 $\mu$s is used for small field sizes and 0 $\mu$s is used for large field sizes.

\section{Stabilization of qubit frequency setpoint}

\begin{figure}[t]
	\begin{center}
		\includegraphics[width=0.5\linewidth]{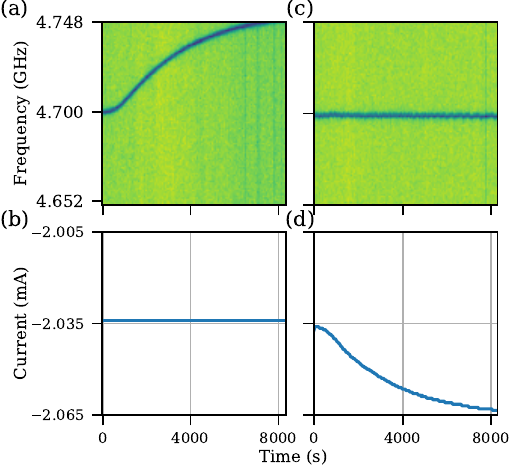}
	\end{center}
	\vspace{-0.6cm}
	\caption{\label{fig:BZservo} 
    Characterization of qubit drift under environmental fluctuations without or with the servo. For these data sets, both the initial qubit frequency and the target frequency are set to 4.7 GHz. (a-b) Qubit spectroscopy and DC bias current as a function of time when the servo is turned off. To simulate environmental fluctuations, we turn on and off another device in the fridge to change the base plate temperature. As the other device is turned on, the qubit frequency steadily increases from 4.7 GHz to above 4.748 GHz over a period of 2 hours. (c-d) Qubit spectroscopy and DC bias current as a function of time when the servo is turned on. For over two hours, the servo actively decreases the DC bias current to maintain qubit frequency at 4.7 GHz within tolerance $\delta_{\mathrm{freq}}= 4$ MHz.
    }
\end{figure}

Environmental fluctuations (such as changes in the lab temperature or base-plate temperature in the dilution fridge) cause the qubit frequency to drift over the course of a measurement. 
To stabilize the qubit about the mean qubit frequency $\omega_{q_0}$ required for the synthetic field, an active servo is implemented through the DC bias line. During each measurement, the servo periodically measures the qubit frequency, and applies a correction to the DC bias current of the external magnet to maintain the qubit at $\omega_{q_0}$.

The servo takes in the desired qubit frequency $\omega_{\mathrm{target}}$ as input. At each adjustment step, it measures the current qubit frequency $\omega$ through a standard spectroscopy measurement. The servo computes a required current correction $I_{\mathrm{corr}} = (\omega_{\mathrm{target}} - \omega)/A$, where $A$ is a predetermined scaling factor, and adjusts the DC bias current to $I_{\mathrm{new}} = I_{\mathrm{old}} + I_{\mathrm{corr}}$. A maximum cap of current correction $I_{\mathrm{cap}}$ is set such that at each step the servo adjusts the DC bias current $I_{\mathrm{old}}$ by only $\min\{ I_{\mathrm{corr}},I_{\mathrm{cap}} \}$. The servo will run for a minimum number of adjustment steps $N_{\mathrm{min}}$ and stop once the measurement frequency is within tolerance $\delta_{\mathrm{freq}} < |\omega_{\mathrm{target}} - \omega|$ or when the number of steps reaches a maximum number $N_{\mathrm{max}}$. Fig.~\ref{fig:BZservo} shows qubit spectroscopy data and corresponding DC bias current data when the servo is off [Fig.~\ref{fig:BZservo}(a)-(b)] versus on [Fig.~\ref{fig:BZservo}(c)-(d)]. Because the device is sensitive to heat load from other devices in the fridge, the data is taken while another device in the fridge is turning on. In both cases, the initial qubit frequency $\omega$ and the target frequency $\omega_{\mathrm{target}}$ are set to 4.7~GHz. When the servo is off, the DC bias current $I$ stays constant throughout the period and the qubit frequency gradually drifts away from $\omega_{\mathrm{target}}$. When the servo is on, it actively adjusts DC bias current $I$. The qubit frequency $\omega$ remains stable around $\omega_{\mathrm{target}} = 4.7$ GHz with a tolerance of $\delta_{\mathrm{freq}}= 4$ MHz.  

\section{Qubit Timescales}\label{appendix:qubit_timescales}
\begin{figure*}
    \centering
    \includegraphics[width=\linewidth]{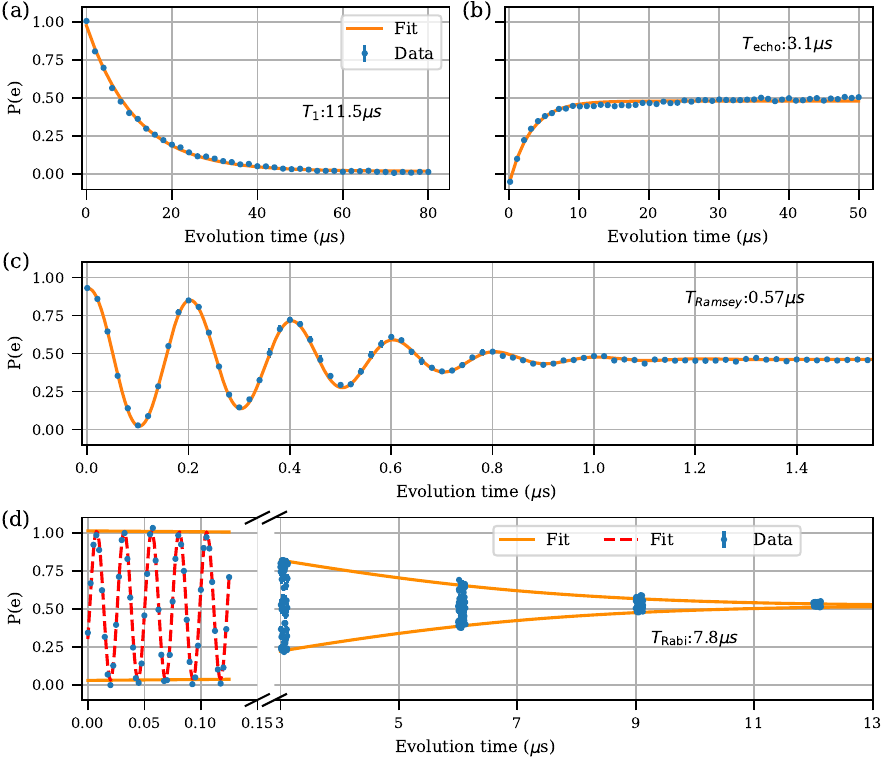}
    \caption{Qubit time scales at 4.7 GHz. All traces are acquired as the average of 10 traces with 10,000 averaged measurements per time point each. The statistical error bars are computed as the standard error of the mean between the traces and are smaller than the markers. (a)~Excited state population as a function of delay time after qubit excitation. $T_1=11.5~\mu$s is extracted from an exponential fit to the data. (b)~Excited state population after a standard Hahn echo sequence. $T_{\mathrm{echo}}=3.1~\mu$s is extracted from an exponential fit to the data. (c)~Ramsey fringes of the qubit population. Due to the short lifetime, a virtual rotation of the second $\pi$/2-pulse was used instead of using a detuned pulse. $T_{\mathrm{Ramsey}}$ is extracted from a Gaussian fit to the envelope of the data. (d)~Rabi oscillations for a 40 MHz drive strength. $T_{\mathrm{Rabi}}$ is extracted from a Gaussian fit to the envelope of the oscillations.}
    \label{fig:Time scales at 4.7GHz}
\end{figure*}

Qubit decoherence-timescale characterization for the qubit-cavity system is performed with the qubit frequency at 4.7 GHz, and shown in Fig.~\ref{fig:Time scales at 4.7GHz}.  Qubit and cavity frequencies, linewidths, and coupling strengths for the system are listed in Table~\ref{tab:qubit_params}. The qubit decay time $T_1 = 11.5\ \mu$s is extracted from an exponential decay fit to the data, shown in Fig.~\ref{fig:Time scales at 4.7GHz}(a). We perform a Ramsey-style measurement with two consecutive $\pi/2$ pulses to extract the pure dephasing time $T_{\phi} = 570$~ns, shown in Fig.~\ref{fig:Time scales at 4.7GHz}(b). To maintain a large separation of scales between the oscillatory and decaying time constants of the Ramsey signal while preserving the signal contrast, we apply a phase rotation to the second $\pi/2$ pulse. Measuring a tunable qubit away from its sweetspot introduces low-frequency noise. Therefore, the pure dephasing time $T_{\phi} = 570$~ns is extracted by fitting the data with a Gaussian decay envelope \cite{quantum_engineer_guide} instead of a standard exponential decay. We perform Hahn echo measurements and extract $T_{\mathrm{echo}} = 3\ \mu$s from an exponential fit, see Fig.~\ref{fig:Time scales at 4.7GHz}(c). All data is taken with 100,000 shots by averaging ten acquisitions each with 10,000 shots per time point.

In addition, we extract the decay time of a 40 MHz Rabi oscillation and find $T_{\mathrm{Rabi}} = 7.8\ \mu$s, shown in Fig.~\ref{fig:Time scales at 4.7GHz}(d). The data is taken by varying the hold time of an on-resonance $x$-drive and is fit with a Gaussian envelope to extract the decay parameter. At high Rabi oscillation frequency, the measurement is sensitive to accumulated phase noise and fluctuations in qubit frequency. We therefore randomize the hold time to transfer phase noise into amplitude noise. Rabi data is taken with 50,000 shots from ten acquisitions with 5,000 shots per time point, and use a Gaussian fit instead of standard exponential to the data. The Rabi oscillations do not quite follow textbook form, likely due to higher levels of the transmon or frequency-dependent decay rates (e.g. due to two-level systems in the environment). Note that the oscillations are not fully symmetric and the probability of $\ket{e}$ does not asymptote to 0.5. Nevertheless, the Gaussian-envelope fit captures the right  time constant, and this time constant $T_{\mathrm{Rabi}}$ is distinctly shorter than $T_d$. 

\section{Device Fabrication and Parameters}
\begin{table}[t]
    \centering
    \begin{tabular}{|c|c|c|}
    \hline
    Parameter & Symbol & Value \\
       \hline \hline
       Qubit g-e frequency & $\omega_q$/(2$\pi$)  & 3.9-7.4 GHz\\
       \hline
       Qubit anharmonicity & $\alpha$/(2$\pi$)    & 240 MHz \\
       \hline
       Qubit-boost cavity coupling & $g_m$/(2$\pi$)       & 13 MHz\\
       \hline
       Qubit-readout cavity coupling & $g_r$/(2$\pi$)       & 90 MHz\\
       \hline 
       Qubit decay rate & $\Gamma_q$/(2$\pi$) & 13.9 kHz \\
       \hline \hline
       Readout cavity frequency & $\omega_r$/(2$\pi$)  & 7.492 GHz\\
       \hline
       Readout cavity linewidth & $\kappa_r$/(2$\pi$)  & 350 kHz\\
       \hline \hline 
       Boost cavity frequency & $\omega_m$/(2$\pi$)  & 5.04 GHz\\
       \hline
       Boost cavity linewidth & $\kappa_m$/(2$\pi$)  & 84 kHz\\
       \hline
    \end{tabular}
    \caption{Device Parameters. All parameters are acquired for the qubit biased at 4.7 GHz.
    $\Gamma_q$ is defined as $1/T_1$, the population decay rate of the qubit from the excited state without Purcell loss into the boost cavity.
    }
    \label{tab:qubit_params}
\end{table}
The device was fabricated on a 530 $\mu$m C-plane sapphire substrate from CrysTec with a 200 nm superconducting tantalum film deposited by StarCyo. The wafers are diced into 7$\times$7 mm chips which are cleaned in PG Remover 1165 before rinsing with acetone and isopropyl alcohol (three solvent clean). All chip features other than the Josephson junctions and leads connecting to the capacitor pads are defined using photolithography followed by a wet etch. The photolithography is performed with a Heidelberg MLA 150 direct write system on AZ1518 resist and developed with CD-26 MIF developer. The pattern is then transferred to the tantalum using a tantalum 111 wet etchant from Transene. Before performing e-beam lithography for the junctions, the chips are cleaned with the three solvent clean followed by a TAMI clean (toluene, acetone, methanol, isopropanol) and finished with a Piranha clean to remove organic residue. For the e-beam lithography, the sample is prepared with a two layer resist stack of MMA (MMA(8.5)MAA EL 13) and PMMA (950PMMA A 3) capped off with an aluminum anti-charging layer. E-beam lithography is performed on an Elionix system (ELS G100) with 0.5 nA write current. Finally, the sample is developed in a MIBK:IPA (1:3) mixture before double angle evaporation in a Plassys MEB550S system. The junctions consist of a 30 nm Al layer followed by oxidation at 2.33 mBar and another 50 nm Al layer deposited on top. The final oxidation occurs at 40 mBar for 10 minutes to encapsulate the deposited aluminum with a smooth oxide. The samples are placed in hot 1165 for a few hours for lift off before a final rinse in acetone and IPA. 
\section{Wiring Diagram}
\begin{figure}
    \centering
    \includegraphics[width=\linewidth]{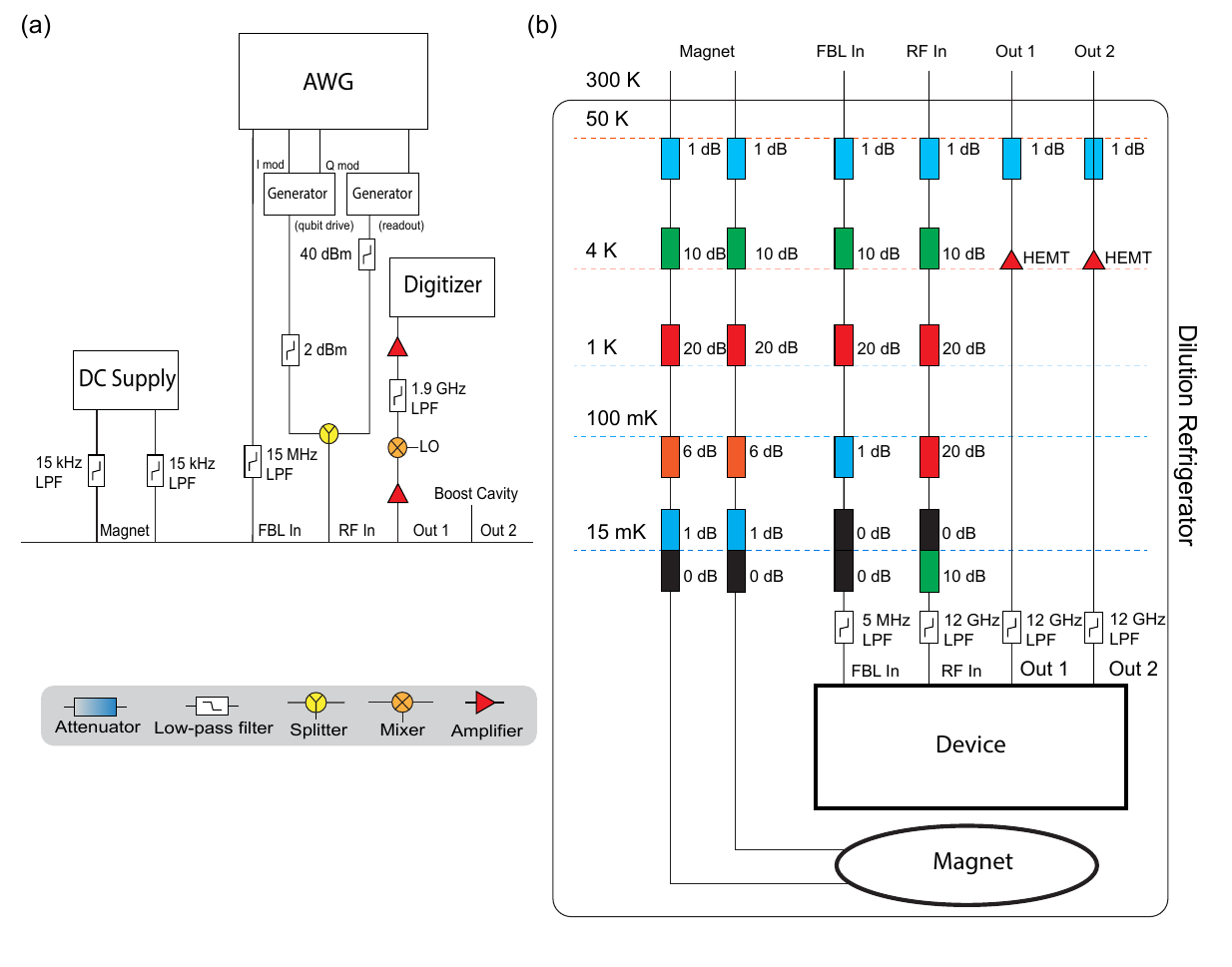}
    \caption{Experimental wiring diagram. (a) Signal-generation and readout hardware. (b) Fridge wiring diagram.}
    \label{fig:wiring_diagram}
\end{figure}

The full wiring diagram for the attenuation and filter configuration in the dilution refrigerator is shown in Fig.~\ref{fig:wiring_diagram}, along with the microwave generation and readout scheme. The qubit frequency is controlled via an external magnet driven with a Keithley 2400 current supply with 15 kHz low pass filters to provide a stable DC set point on the qubit. For the fast frequency modulation required by the protocol, the qubit frequency is varied using the on-chip flux bias line with a 5 MHz low pass filter to reduce the noise bandwidth reaching the qubit. 

\end{document}